\documentclass[final,authoryear,11pt]{elsarticle} %use for preparation in Shell - saves paper!

\usepackage{graphicx,amssymb,amsfonts,amsthm}
\usepackage[tbtags]{amsmath}
\usepackage{natbib}
\usepackage[utf8]{inputenc}
\usepackage[british]{babel}
\usepackage[mathscr]{eucal}
\usepackage[toc,page]{appendix}

\pdfoutput=1

\journal{Atmospheric Environment}

\providecommand{\C}{\mathcal{C}} %source characteristics other than strength
\providecommand{\D}{\mathcal{D}} %extended data
\providecommand{\R}{\mathbb{R}} %real numbers
\providecommand{\W}{\mathcal{W}} %wind field
\providecommand{\vect}[1]{\mathbf{#1}}
\providecommand{\myb}{\vect{b}} %background
\providecommand{\s}{\vect{s}} %source strengths
\providecommand{\w}{\vect{w}} %vector wind field
\providecommand{\x}{\vect{x}} %trajectory
\providecommand{\y}{\vect{y}} %measured concentration
\providecommand{\z}{\vect{z}} %source grid locations
\providecommand{\bta}{\boldsymbol{\beta}} %background parameters
\providecommand{\eps}{\boldsymbol{\epsilon}} %measurement error
\providecommand{\ph}{\boldsymbol{\phi}} %auxilliary rvs in RJMCMC step
\providecommand{\tht}{\boldsymbol{\theta}} %auxiliary source parameters
\providecommand{\thtkpp}{\boldsymbol{\theta}_{\kappa}} %auxiliary source parameters
\providecommand{\thtnotkpp}{\boldsymbol{\theta}_{\overline{\kappa}}} %auxiliary source parameters
\providecommand{\A}{\vect{A}} %source strengths

\providecommand{\J}{\vect{J}} %source strengths
\renewcommand{\P}{\vect{P}} %source strengths
\providecommand{\Q}{\vect{Q}} %source sparsity
\providecommand{\U}{\vect{U}} %source strengths

\providecommand{\diff}[2]{\frac{\partial #1}{\partial #2}} %partial derivative (2 arguments)
\providecommand{\by}[2]{#1 \times #2}
\newcommand{\mycite}[1]{ (\cite{#1})}
\begin{document}

\begin{frontmatter}


\title{Locating and quantifying gas emission sources using remotely obtained concentration data}
\author{Bill Hirst}
\address{Shell Global Solutions International B.V., The Netherlands}
\cortext[cor1]{bill.hirst@shell.com, Tel:+31 6 5512 3526, Fax:+31 70 447 2393}
\author{Philip Jonathan}
\cortext[cor2]{philip.jonathan@shell.com, Tel:+44 151 272 5421, Fax:+44 151 +44 5384}
\address{Shell Technology Centre Thornton, UK.}
\author{Fernando Gonz\'alez del Cueto}
\address{Shell International Exploration and Production, U.S.A.}
\author{David Randell}
\address{Shell Technology Centre Thornton, UK.}
\author{ Oliver Kosut}
\address{Laboratory for Information and Decision Systems, Massachusetts Institute of Technology, USA.}

\begin{abstract}
We describe a method for detecting, locating and quantifying sources of gas emissions to the atmosphere using remotely obtained gas concentration data; the method is applicable to gases of environmental concern. We demonstrate its performance using methane data collected from aircraft.   Atmospheric point concentration measurements are modelled as the sum of a spatially and temporally smooth atmospheric background concentration, augmented by concentrations due to local sources.  We model source emission rates with a Gaussian mixture model and use a Markov random field to represent the atmospheric background concentration component of the measurements. A Gaussian plume atmospheric eddy dispersion model represents gas dispersion between sources and measurement locations.  Initial point estimates of background concentrations and source emission rates are obtained using mixed $\ell_2$-$\ell_1$ optimisation over a discretised grid of potential source locations. Subsequent reversible jump Markov chain Monte Carlo inference provides estimated values and uncertainties for the number, emission rates and locations of sources unconstrained by a grid. Source area, atmospheric background concentrations and other model parameters are also estimated. We investigate the performance of the approach first using a synthetic problem, then apply the method to real data collected from an aircraft flying over: a $1600 \text{km}^2$ area containing two landfills, then a $225 \text{km}^2$ area containing a gas flare stack.
\end{abstract}

\end{frontmatter}

%%%%%%%%%%%%%%%%%%%%%%%%%%%%%%%%%%%%%%%%%%%%%%%%%%%%%%%%%%%%
%\vspace{250pt}
%\pagebreak
%\tableofcontents
%%%%%%%%%%%%%%%%%%%%%%%%%%%%%%%%%%%%%%%%%%%%%%%%%%%%%%%%%%%%
%Outstanding work
%***
%Nothing
%%%%%%%%%%%%%%%%%%%%%%%%%%%%%%%%%%%%%%%%%%%%%%%%%%%%%%%%%%%%

%%%%%%%%%%%%%%%%%%%%%%%%%%%%%%%%%%%%%%%%%%%%%%%%%%%%%%%%%%%%
%Introduction
\section{Introduction} \label{Int}
There is growing interest in developing methods for detecting and locating sources of gas emissions into the atmosphere.   Greenhouse gases are of intense interest (e.g. \cite{Chen06, Shakhova10}). Other applications include monitoring toxic gas emissions, locating explosives from their volatile emissions (e.g. \cite{Bhattacharjee08}]), mapping naturally occurring gas seeps for oil and gas exploration (e.g. \cite{Hirst04}), identifying sources of nuisance odours, and even understanding how moths are able to find mates by detecting pheromones at concentrations corresponding to individual molecules (e.g. \cite{Vergassola07}). For greenhouse gases and oil and gas exploration the goal is to locate sources and quantify emission rates.  For explosives, nuisance odours and moths, locating the source is sufficient!

In this paper we concentrate on the task of detecting, locating and quantifying the emission rates of sources of a single gas species of interest.  While the method we have developed is broadly applicable to any gas or passively transported, detectable atmospheric component (e.g. aerosols, radon, smoke, dust, viruses) we concentrate on methane emissions using data acquired during development and testing of an airborne system for mapping natural gas seeps for use in hydrocarbon exploration over areas typically up to 5000 $\text{km}^2$ per flight.  We apply the method to measurements from test flights around two modern Canadian landfills and a flare stack within a modern natural gas processing facility in North Africa, collected at ranges of up to $12$km downwind of sources. We make inferences about source emission rates using gas concentration measurements. We measure concentrations by volume and use our understanding of gas dispersion to relate these to source emission rates expressed as $\text{m}^3\text{s}^{-1}$ of pure gas per source. For area sources emission rates can also be expressed as mass flux; i.e. mass emission rate per unit time per unit area.  Critical to achieving this goal is estimating the level of atmospheric background concentration, so we can identify the additional concentration over and above background, attributable to the local source(s) of interest. The novelty of the current work lies in the tailored application of a combination of standard statistical modelling components and inference tools for inversion in remote sensing.

The extensive literature on inversion and the related field of compressive sensing (e.g. \cite{Donoho2006}) includes contributions from the statistics, applied mathematics, electrical engineering and physics communities. \cite{Sambridge2002} reviews the development and application of Monte Carlo methods for inverse problems in the Earth sciences. \cite{Rao2007} reviews source estimation methods for atmospheric releases of toxic agents, including forward modelling (potentially using Bayesian inference) and backward transport modelling, emphasising the need to assess uncertainties in characterisation of sources using atmospheric transport and dispersion models. \cite{SncEA2008} discusses early detection of the location and size of a contaminant release into the atmosphere from a network of environmental sensors using a Gaussian plume forward model with stochastic parameters and Bayesian inference using Markov chain Monte Carlo (MCMC). It is noted that the main distinguishing feature of Bayesian inference as opposed to optimisation (e.g. \cite{HrsEA2007}) for source estimation is that the former estimates probability distributions for parameters of interest and quantifies the uncertainty in the estimated parameter, whereas the latter provides point estimates for the parameters of interest through optimising an objective function. \cite{KtsEA2007} note that determining the source of an emission from the limited information provided by a finite and noisy set of concentration measurements obtained from real-time sensors is an ill-posed inverse problem. They show that solving the adjoint advection-diffusion equation just once per detector location allows efficient forward model estimation and Bayesian inference using MCMC. \cite{LngEA2010} uses a genetic algorithm to find the combination of source location, height and strength, surface wind direction and speed, and time of release that produces a concentration field that best matches sensor observations. A rationale is developed to specify the minimum number of sensors necessary to estimate the source term and to obtain the relevant wind information to a given precision. \cite{RddEA2012} presents an inverse modelling technique to estimate source strength and location, together with the uncertainty in those estimates, using a limited number of measurements from a sensor network. Experimental design aspects are addressed, including the optimal number and configuration of sensors for a given measurement campaign, and the minimum period of observation for source detection with confidence. The need to relate uncertainty in estimated source properties to those of the input data is emphasised. \cite{GSzEA2011} introduces a Bayesian multiple change-point model for monitoring of air quality standards by pollutants such as nitrogen oxides and carbon monoxide. Change-points are identified and the rate of occurrence of air quality threshold exceedence estimated using a reversible-jump MCMC approach.

The layout of the paper is as follows. Motivating landfill and flare stack applications are first described in section \ref{Dat}. Section \ref{Mdl} outlines the modelling procedure and illustrates it in application to a synthetic problem. Application of the method to the landfill and flare stack measurements is described in section \ref{App}. Section \ref{Dsc} provides a discussion of findings and suggestions for future development. Modelling details are relegated to three appendices, describing background modelling (\ref{App:Bck}), initial parameter estimation using mixed optimisation (\ref{App:IntPrmEst}) and mixture modelling (\ref{App:MxtMdl}).
%%%%%%%%%%%%%%%%%%%%%%%%%%%%%%%%%%%%%%%%%%%%%%%%%%%%%%%%%%%%

%%%%%%%%%%%%%%%%%%%%%%%%%%%%%%%%%%%%%%%%%%%%%%%%%%%%%%%%%%%%
%Data
\section{Data} \label{Dat}
We used an ultra-sensitive, high precision methane gas sensor, mounted in an aircraft, to measure a continuous stream of air from the leading edge of a wing -- well away from any fuel/lubricant or engine exhaust fumes. The sensor continuously measures gas concentration and passes data to the aircraft's data logging system together with GPS, radar altitude, barometric pressure, air temperature and wind velocity data, and several system control parameter values.  The sensor delivers better than $1$ ppb (part per billion by volume) precision concentration data with a response time of approximately $1$ second. Flight data are subsequently merged with specialist meteorological data, including additional wind, atmospheric boundary layer depth and auxiliary data: such as the air sample transit time from sample inlet to sensor measurement chamber.

The data sets presented here are atypical of surveys aimed at hydrocarbon exploration, in that we know the source locations within in the survey areas. Consequently, these data sets provide a valuable test of our measurement and analysis procedures. Source related concentrations are up to two orders of magnitude greater than those typically encountered in natural seepage surveys.  Landfill measurements provide a direct test of the total system performance; flare stack measurements provide a means of determining the gas sample transit time in flight at operational speeds and allow us to probe the 3 dimensional structure of the gas plume. This is essential to correctly assign concentration data to measurement locations.

\subsection{Landfills} \label{Dat:LndFll}

Atmospheric methane is responsible for about one third of the global warming effect of $CO_2$ despite $CO_2$ concentrations being  approximately $220$ times greater than those of methane.  There are strong incentives to reduce methane emissions to the atmosphere.  Landfills are prodigious sources of anthropogenic methane; about $25\%$ of United States' methane emissions are from landfills,\citep{EPA09}.

The global average atmospheric methane concentration is approximately $1,820$ppb, increasing at approximately $8$ppb per annum \citep{Dlu09}. Local atmospheric background methane concentration can vary by approximately $20$ ppb during daytime due to changes in Atmospheric Boundary Layer (ABL) depth. This layer of the atmosphere effectively contains all ground level emissions for that day. Its growth is driven by solar heating of the ground.  In effect, ground level emissions accumulate near the ground during the night when the ABL is thin and are diluted into the growing volume of the ABL during daytime.  It is important to account for the associated changes in local background concentration, so as to more precisely determine that portion of measured concentration attributable to the local sources of interest: since source related concentrations can be comparable to the much longer term changes in background.

The top of the ABL constitutes a ``ceiling'' on vertical transport of gases from the ground, and reduces the rate of dilution with downwind distance from the source.  This effect must be included in the gas dispersion model used to relate measured concentration to source strength. It can be modelled as the sum of multiple reflections from the ABL ``ceiling'' and ground surface. At longer ranges such as those prevailing here, there will be multiple bounce terms successively from the ABL ``ceiling'' and the ground surface, until the air within the ABL is well mixed (\cite{Gifford76}). The aircraft must be within the ABL if it is to detect concentrations from ground level sources.

\begin{figure}[ht!]
  \center\includegraphics[width=1\textwidth]{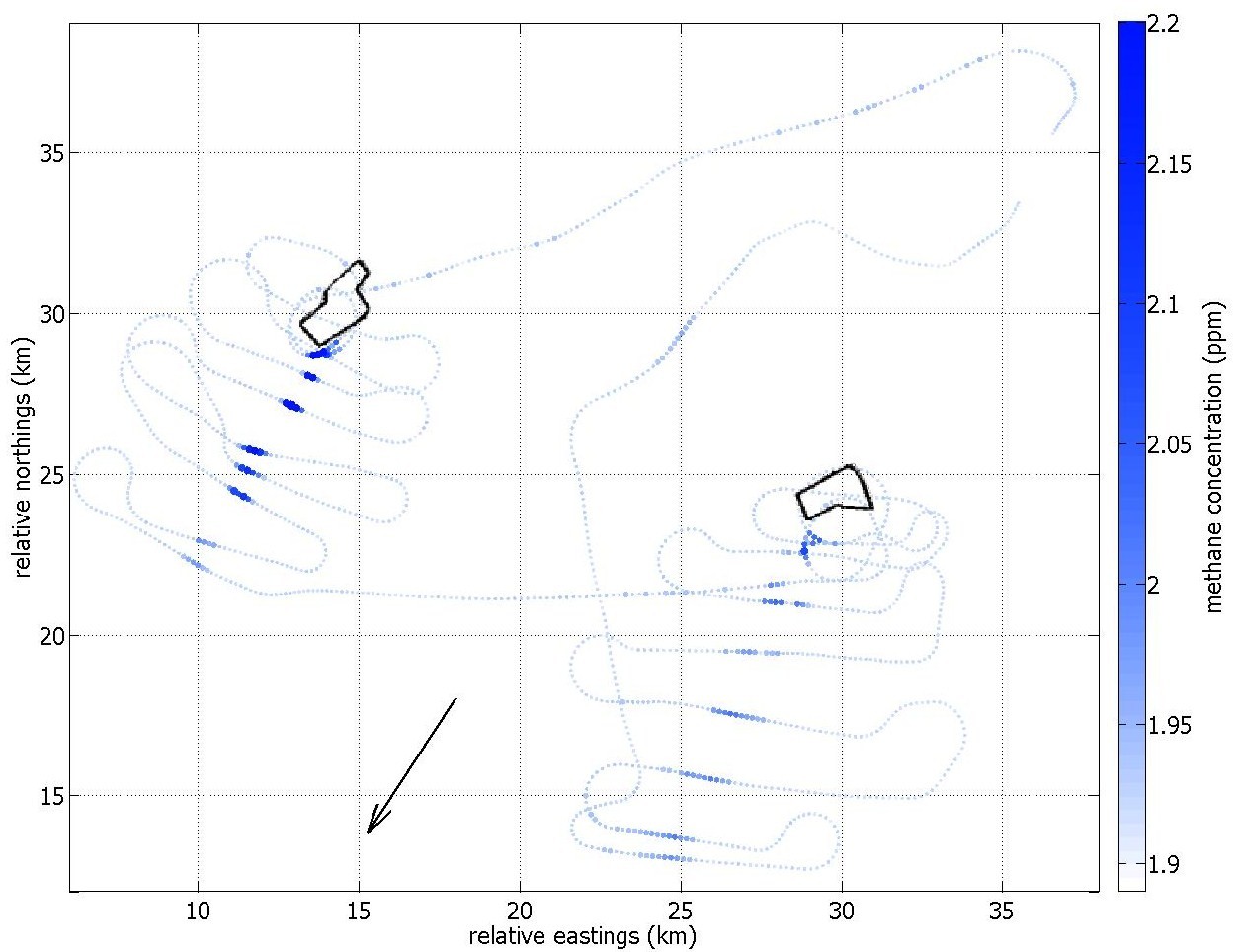}\\
  \caption{Flight track around and in vicinities of two landfills. Blue marker size and colour saturation indicate strength and location of measured methane concentrations. Arrow shows average direction of predicted air movement during flight. Polygons show perimeters of methane--emitting landfill areas. Dimensions in $km$. Aircraft takes off in North East corner and flies over Westerly landfill first.}
  \label{LndFll_SrcMapCnC}
\end{figure}

The data presented here correspond to measurements at the aircraft, which flies at approximately $200$m AGL (above ground level) at a speed of approximately $50\text{ms}^{-1}$; ABL depth is approximately $400$m. The area of interest is $40\text{km} \times 40\text{km}$ and the flight time approximately $80$ minutes.  Figure \ref{LndFll_SrcMapCnC} shows the flight track of the aircraft around each site, tracing a loose serpentine pattern downwind of each landfill.  Initial average wind speed and direction (based on measurements near the site and model data provided for multiple altitudes by the UK Meteorological Office, UKMO \cite{UKMO05} ) are about $6.5\text{ms}^{-1}$ and $033^o$ degrees meteorological. (Wind direction is defined as the direction \emph{from which} the wind blows, in a clockwise sense, with North as $0^o$. In this case, the wind direction is approximately North-Easterly, i.e. air moves approximately to the South West.).

\begin{figure}[ht!]
  \center\includegraphics[width=0.9\textwidth]{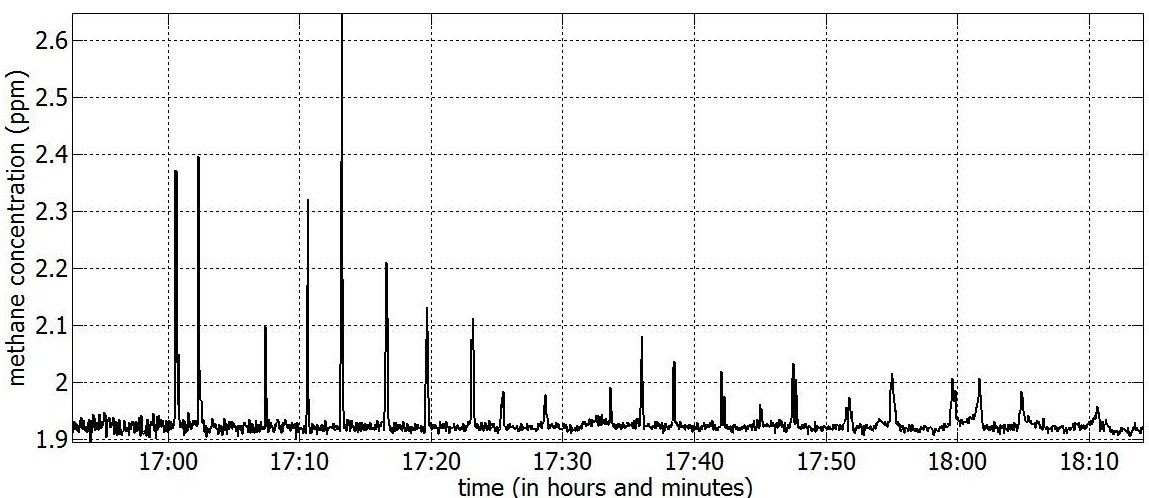}\\
  \caption{Methane concentrations along landfill flight path as a function of time.}
  \label{LndFll_CnC}
\end{figure}

Figure~\ref{LndFll_SrcMapCnC} shows the flight trajectory around and in the vicinity of the two landfills; Figure~\ref{LndFll_CnC} shows measured methane concentrations in time. Inspection of the concentration trace along the trajectory in Figure~\ref{LndFll_SrcMapCnC} indicates the wind direction is approximately constant for the period of the flight.

\subsection{Flare stack} \label{Dat:FlrStc}

The flare stack flight comprises $8$ separate multi--looped circuits of the flare at altitudes from $150$m to $350$m AGL; ABL depth is greater than $1500$m.  Each circuit intersects the plume three times at different angles and ranges to probe the $3$-D structure of the dispersion plume.  Additionally measurements are used to determine the air sample transit time to provide consistent plume delineation.  The flare stack is $50$m high, situated within a recently completed natural gas processing plant at a coastal location, where winds are variable, probably burning methane and light hydrocarbons. Photographs of the flare stack show it to be a clean yellow flame, which suggests it is burning at high efficiency \citep{Kea00}. Methane content of a few percent is expected in the residual unburnt gas; this is due to thermal decomposition and incomplete combustion of the fuel-rich hydrocarbon feed \citep{Pek05}.

\begin{figure}[ht!]
  \center\includegraphics[width=0.85\textwidth]{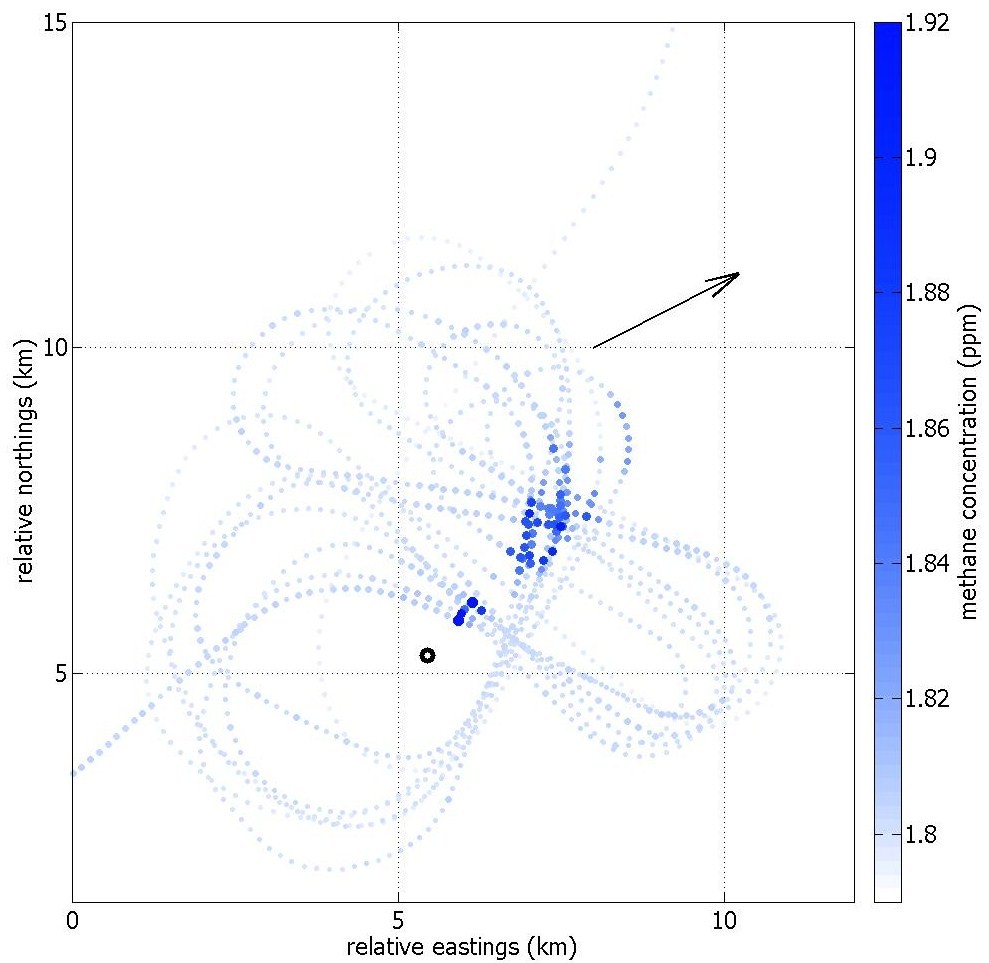}\\
  \caption{Flight track around and in vicinity of the flare stack. Blue marker size and colour saturation indicate strength and location of measured methane concentrations. Arrow shows average direction of predicted air movement during flight. Black annulus indicates location of flare. The visible discrepancy in alignment of significant concentrations and flare stack with respect to arrow indicates error in predicted direction of air movement. Aircraft enters from the SW corner and leaves NE corner.}
  \label{FlrStc_SrcMapCnC}
\end{figure}

\begin{figure}[ht!]
  \center\includegraphics[width=0.9\textwidth]{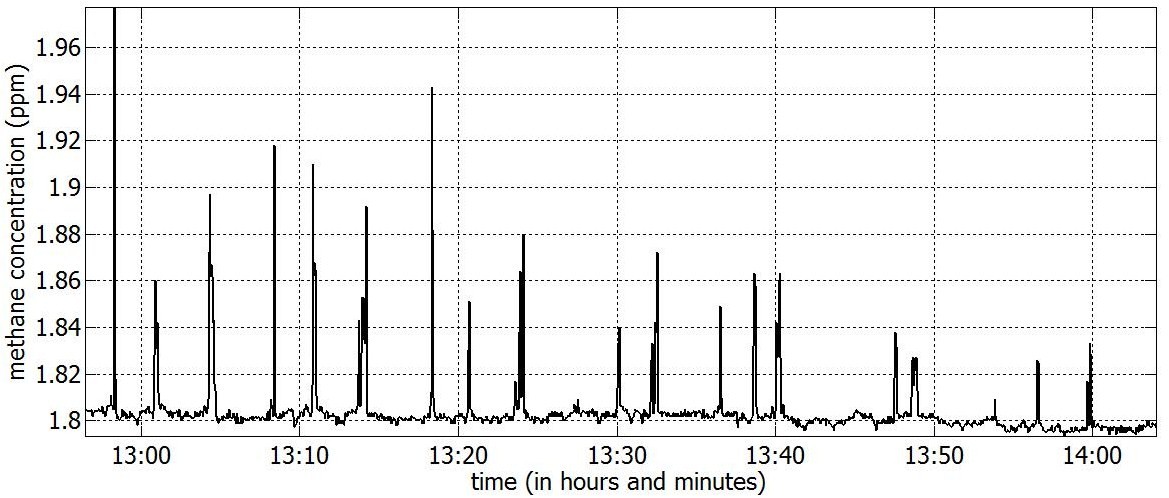}\\
  \caption{Methane concentrations along flare stack flight path as a function of time.}
  \label{FlrStc_CnC}
\end{figure}

Initial UKMO model--based predicted average wind speed and direction are $11\text{ms}^{-1}$ at $-243^o$. Figure \ref{FlrStc_SrcMapCnC} shows the flight track around and in the vicinity of the flare stack; Figure \ref{FlrStc_CnC} shows measured methane concentrations against time. Inspection of the concentration trace along the track in Figure \ref{FlrStc_SrcMapCnC} suggests that the UKMO predicted wind direction is inaccurate for the portion of the flight: which is late in the afternoon and situated over the coastline.

%%%%%%%%%%%%%%%%%%%%%%%%%%%%%%%%%%%%%%%%%%%%%%%%%%%%%%%%%%%%

%%%%%%%%%%%%%%%%%%%%%%%%%%%%%%%%%%%%%%%%%%%%%%%%%%%%%%%%%%%%
%Model
\section{Model} \label{Mdl}
\subsection{Model form} \label{Mdl:Frm}

We seek to locate and quantify emission rates of methane sources given $n$ observed atmospheric concentration measurements $\y=\{y_i\}_{i=1}^n$ along airborne trajectory $\x=\{\x_i\}_{i=1}^n$. The overwhelming majority of concentration measured is the atmospheric background contribution of about $1800$ppb with an additional $0-100$ppb (typically) attributable to local ground level sources. Background level varies in space and time. Poor estimation of background concentration disproportionately impacts estimation of concentration attributable to local sources. Hence we prefer to jointly estimate background and source contributions. $\y$ is modelled as the sum of a slowly varying background $\myb=\{b_i\}_{i=1}^n$ along the trajectory and contributions due to a distributed group of $m$ sources at ground level locations $\z=\{\z_j\}_{j=1}^m$ with emission rates $\s=\{s_j\}_{j=1}^m$ and auxiliary characteristics $\C=\{\C_j\}_{j=1}^m$.  Measurements along the trajectory are assumed to be made with independent identically-distributed additive Gaussian errors $\eps=\{\epsilon_i\}_{i=1}^n$, with $\epsilon_i \sim N(0,\sigma_{\epsilon}^2)$. Steady-state gas transport between a source of unit emission rate at location $\z_j$ and measurement location $\x_i$ in wind field $\W$ is given by coupling function $a_{ij}=a(\x_i,\z_j,\C_j;\W)$. With $A=\{a_{ij}\}_{i=1,j=1}^{n,m}$, we adopt the model:
\begin{equation} \label{E:MdlY}
\y = \A \s + \myb + \eps
\end{equation}
Model mis-specification can be diagnosed by analysis of residuals, e.g. the elements of $\eps$ in equation \ref{E:MdlY} should represent a random sample from a normal distribution with mean zero and constant variance.

\subsection*{Plume model}

\begin{figure}[ht!]
  \center\includegraphics[width=1\textwidth]{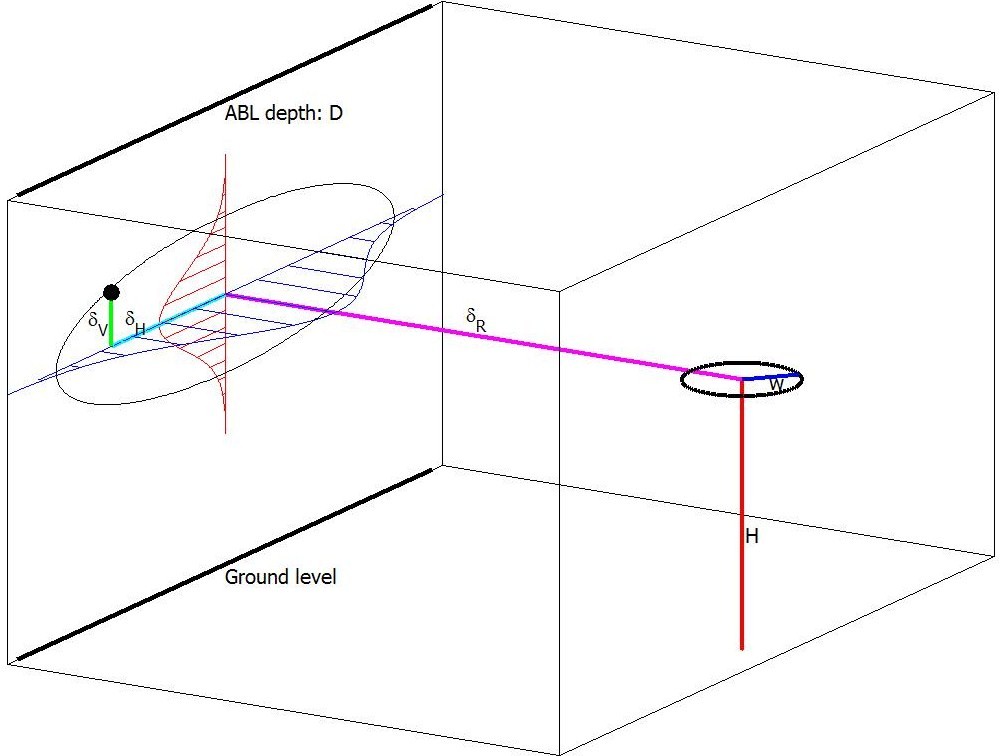}\\
  \caption{  Illustration of plume model parameters. Red line: source height $H$. Magenta line: downwind distance $\delta_R$ of measurement location relative to the source. Cyan line: horizontal offset $\delta_H$. Green line: vertical offset $\delta_V$. Blue line: source half width $w$. Thick black horizontal lines perpendicular to wind direction represent ground level and top of ABL at height $D$. The current plume model allows for reflections from ground and ABL ceiling. In this figure $\delta_R$ is too short for reflections to be applicable. The hatched blue area represents the marginal variation with $\delta_H$ of the value of $a$ for $\delta_V=0$ and fixed $\delta_R$. The hatched red area represents the marginal variation with $\delta_V$ of the value of $a$ for fixed $\delta_H$  and $\delta_R$. The locus of the black dot (corresponding to a contour of constant $a$ for fixed $\delta_R$) is drawn as a black closed curve. }
  \label{fig:Plume}
\end{figure}

The wind field $\W$ is described by wind vector $\U$ at measurement location $x$, and horizontal and vertical plume opening angles $\gamma_H$ and $\gamma_V$ respectively. We approximate coupling $a(x,z,w;\U, \gamma_H, \gamma_V)$ between a unit source of half width $w$ at location $z$ and measurement location $x$ in wind field $\W$ by a Gaussian plume model. The measurement location relative to the plume is expressed in terms of downwind distance $\delta_R$, and horizontal and vertical offsets $\delta_H$ and $\delta_V$ of measurement location with respect to wind vector. $\sigma_H = \delta_R \tan(\gamma_H)+w$ and $\sigma_V = \delta_R \tan(\gamma_V)$ play the role of plume ``standard deviations'' in horizontal and vertical directions respectively.
\begin{align}
a(x,z,w;\U, \gamma_H, \gamma_V) =& \frac{1}{2\pi |\U| \sigma_H \sigma_V} \exp\left\{-\frac{\delta_H^2}{2 \sigma_H^2}\right\} \times  \left\{ \quad\exp\Big\{-\frac{(\delta_V-H)^2}{2 \sigma_V^2}\Big\}+\exp\Big\{-\frac{(\delta_V+H)^2}{2 \sigma_V^2}\Big\}\right. \nonumber \\ & \quad \quad \quad +  \left. \exp\Big\{-\frac{(2D-\delta_V-H)^2}{2 \sigma_V^2}\Big\} +\exp\Big\{-\frac{(2D-\delta_V+H)^2}{2 \sigma_V^2}\Big\} \quad \right\}\label{E:Plm}
\end{align}
where $D$ is the maximum altitude of the atmospheric boundary layer above ground level (referred to as the atmospheric boundary layer depth). The sum of four exponential terms represents plume reflections in the ground and at the interface between atmospheric boundary layer and free atmosphere above it at altitude $D$. Values of $\U$, $D$, $\gamma_H$ and $\gamma_V$ are obtained directly from wind field data supplied by UKMO. The plume model parameters are illustrated in Figure \ref{fig:Plume}. Typical plume characteristics are shown and discussed in Section \ref{App:IllAnl}.

\subsection*{Background model} \label{Mdl:BckMdl}

Well-mixed background gas concentration $\myb$ along the trajectory is assumed to be positive and smoothly--varying spatially and temporally. We assume it can be represented by an appropriate set of $r$-dimensional smooth spatio-temporal basis functions:
\begin{equation} \label{E:MdlB}
\myb=\P\bta
\end{equation}
for an $n \times r$ matrix $\P$ of bases, where $\bta=\{\beta_k\}_{k=1}^r$ are parameters to be estimated. As a default approach, we assume $\P$ to be the identity matrix and adopt a Gaussian Markov random field model for $\bta (\equiv \myb)$. In this case, $\bta$ is a random vector with prior probability density function:
\begin{equation} \label{E:PrpBta}
f(\bta) \propto \exp\{ - \tfrac{\mu}{2} (\bta-\bta_0)^T (\J_{\bta}) (\bta-\bta_0) \}
\end{equation}
with pre-specified precision matrix $\J_{\bta}$ and tuning parameter $\mu$. Due to the random field's conditional independence structure, $\J_{\bta}$ is guaranteed to be sparse, allowing efficient parameter estimation. As outlined in Appendix \ref{App:Bck}, we specify the precision matrix such that pairs of locations aligned with the wind direction have higher dependence. We also consider other representations of the form $\myb=\P\bta$, e.g. polynomial or spline bases for which $r \ll n$ (see Appendix \ref{App:Bck}).

\subsection{Initial parameter estimation} \label{Mdl:IntPrmEst}

Given measured concentrations $\y$ on trajectory $\x$ and associated wind field data $(\{\U(\x_i)\}_{i=1}^n, \gamma_H, \gamma_V)$, initial point estimates for source emission rates $\s$ and background parameters $\bta$ are obtained as follows. We assume a Laplace prior for $\s$ with pre-specified precision $\Q$ and tuning parameter $\lambda$:
\begin{equation} \label{E:PrpS}
f(\s) \propto \exp\{ - \lambda \|\Q \s \|_1 \}
\end{equation}
Since the likelihood corresponding to model \ref{E:MdlY} is:
\begin{equation} \label{E:LklSB}
f(\y | \s,\bta) \propto \exp\{ - \tfrac{1}{2 \sigma_{\epsilon}^2} \|A\s + P\bta -\y \|^2 \} \text{,}
\end{equation}
by applying Bayes theorem, the posterior parameter density becomes:
\begin{equation} \label{E:PstSB}
f(\s,\bta | \y) \propto f(\y | \s,\bta) f(\s) f(\bta)
\end{equation}
In particular the maximum a-posteriori parameters are obtained by maximising $f(\s,\bta | \y)$ with respect to $\s$ and $\bta$, or equivalently by minimising $-\log_e f(\s,\bta | \y)$. In optimisation terms, initial parameter estimation can be stated as:
\begin{equation} \label{E:ObjFnc}
  \mathrm{argmin}_{\s,\bta} \qquad \tfrac{1}{2 \sigma_{\epsilon}^2} \|A\s + P\bta -\y \|^2 + \tfrac{\mu}{2} (\bta-\bta_0)^T J (\bta - \bta_0) + \lambda\|Q\s\|_1 \\
\end{equation}
where terms in $\mu$ and $\lambda$ can be viewed as regularisations that impose background smoothness and source sparsity respectively. We further choose to restrict the domain of source elements such that $\s_j \in [0,s_{\max}]$, and background $\myb_i \in [0,y_i+\tau]$ for tolerance $\tau$. Details are given in Appendix \ref{App:IntPrmEst}.

\subsection{Mixture model} \label{Mdl:MxtMdl}

Full parameter estimation is performed using a mixture modelling approach. We assume that each of $m$ sources can be represented as a two-dimensional Gaussian kernel located at $z_j$ with half width $w_j$ (corresponding to the standard deviation of the Gaussian) and source emission rate $s_j$. Using reversible jump Markov chain Monte Carlo (RJMCMC)  simulation (\cite{Green1995}), we treat $m$ as a random variable, and estimate the joint distribution of $m$ and all other model parameters. We can also make inferences about apparent bias and/or uncertainty in wind field parameters and measurement error. Bias-correction of wind direction proves to be important in some applications. RJMCMC for mixtures of univariate Gaussians was considered by \cite{Richardson1997} and extended to multivariate Gaussian mixture models by \cite{Zhang2004}.

Markov chain Monte Carlo (MCMC) is a simulation procedure for Bayesian inference, which exploits the fact that Markov chains have stationary distributions exhibiting reversible transitions. We seek to create a Markov chain with the posterior distribution $f(\{\z,\w,\s,\bta\} | \y)$ as its stationary distribution using the Metropolis-Hastings algorithm (\cite{Metropolis1953} and \cite{Hastings1970}). Extending the posterior density in \eqref{E:PstSB} to include source locations and half widths, and writing the resulting parameter set $\{\z,\w,\s,\bta\}$ as $\tht$ for brevity, the expression for the posterior parameter density becomes:
\begin{equation} \label{E:PstZWSB}
f(\tht | \y) \propto f(\y | \tht) f(\tht)
\end{equation}
Enumerating the constant of proportionality in \eqref{E:PstZWSB} is generally computationally costly, and hence so is direct sampling from the posterior. Fortunately MCMC (e.g. using Metropolis-Hastings acceptance sampling or similar) circumvents the need to evaluate the constant of proportionality. We proceed by judiciously partitioning the set of model parameters $\tht$ (to exploit problem structure and ensure reasonable MCMC performance), so that dependent parameters appear in the same subset $\thtkpp$ of parameters indexed by $\kappa$ (see, e.g. \cite{GmmLpz06}), and that different sampling techniques (such as Gibbs sampling, see below) can be exploited for different subsets. The conditional posterior distribution of parameter subset $\thtkpp$ becomes:
\begin{equation} \label{E:CndPst}
f(\thtkpp | \y, \thtnotkpp) \propto f(\y | \thtkpp, \thtnotkpp) f(\thtkpp | \thtnotkpp)
\end{equation}
where $\thtnotkpp$ represents the remaining model parameters. The Metropolis-Hastings algorithm then provides a basis for acceptance of a candidate $\{\thtkpp',\thtnotkpp\}$ (with $\thtkpp'$ generated by sampling from a suitable proposal $q(\thtkpp,\thtkpp'|\thtnotkpp)$, such as a multivariate Gaussian centred at $\thtkpp$) given the current Markov chain position $\{\thtkpp,\thtnotkpp\}$ with probability $\alpha(\thtkpp,\thtkpp' |\thtnotkpp)$:
\begin{equation} \label{E:MtrHst1}
\alpha(\thtkpp,\thtkpp' |\thtnotkpp) = \max \left\{1,\  \frac{f(\thtkpp'|\y,\thtnotkpp) q(\thtkpp',\thtkpp| \thtnotkpp)}{f(\thtkpp|\y,\thtnotkpp) q(\thtkpp,\thtkpp'| \thtnotkpp)}\right\}
\end{equation}
Since the conditional posterior appears in both the numerator and denominator of \eqref{E:MtrHst1}, the fact that the conditional posterior is only known in \eqref{E:CndPst} up to a constant of proportionality is of no consequence. Starting from an arbitrary starting point, having allowed for a period of burn-in (to facilitate convergence \citep{GmmLpz06}), the sequence of points $\tht=\{\z,\w,\s,\bta\}$ on the Markov chain will converge to a (dependent) sample from $f(\z,\w,\s,\bta | \y)$. In this way, for a fixed number $m$ of sources, we can estimate the joint posterior distribution of the model parameters.

RJMCMC allows sampling from distributions for which the number of sources $m$ (and hence the total number of model parameters) is not fixed. As explained in Appendix \ref{App:MxtMdl}, the Metropolis-Hastings algorithm can be extended to accommodate ``birth'' of a new source, ``death'' of an existing source, coalescence of neighbouring sources and source division. The Markov chain will therefore explore estimates of $\z$, $\w$ and $\s$ of different dimensions together with $\bta$. Inference can be extended to include estimation of quantities such as the measurement error standard deviation $\sigma_{\epsilon}$, bias of wind vector $\U$ and horizontal and vertical plume opening angles $\gamma_H$ and $\gamma_V$. In the analysis reported in section \ref{App}, $\sigma_{\epsilon}$ (see \eqref{E:MdlY}), and additive wind direction bias (used in \eqref{E:Plm}) are included as model parameters in the Bayesian inference.

When the conditional posterior distribution $f(\thtkpp | \y, \thtnotkpp)$ can be written in closed form, values of $\thtkpp$ can be sampled directly given current values of $\thtnotkpp$. This approach, known as Gibbs sampling (e.g. \cite{Geman1984}) or sampling from full conditionals, avoids the need for acceptance sampling. In the current work, the conditional distribution of background parameters $\bta$ is known in closed form and is amenable to sampling from full conditionals. When full conditionals are not available we use the Metropolis-Hastings algorithm.

The initial optimisation solution (section \ref{Mdl:IntPrmEst}) is sampled to give a suitable starting point to accelerate MCMC convergence to the stationary distribution.
%%%%%%%%%%%%%%%%%%%%%%%%%%%%%%%%%%%%%%%%%%%%%%%%%%%%%%%%%%%%

%%%%%%%%%%%%%%%%%%%%%%%%%%%%%%%%%%%%%%%%%%%%%%%%%%%%%%%%%%%%
%Application
\section{Application} \label{App}
\subsection{Illustrative analysis} \label{App:IllAnl}

\begin{figure}[ht!]
  \center\includegraphics[width=0.97\textwidth]{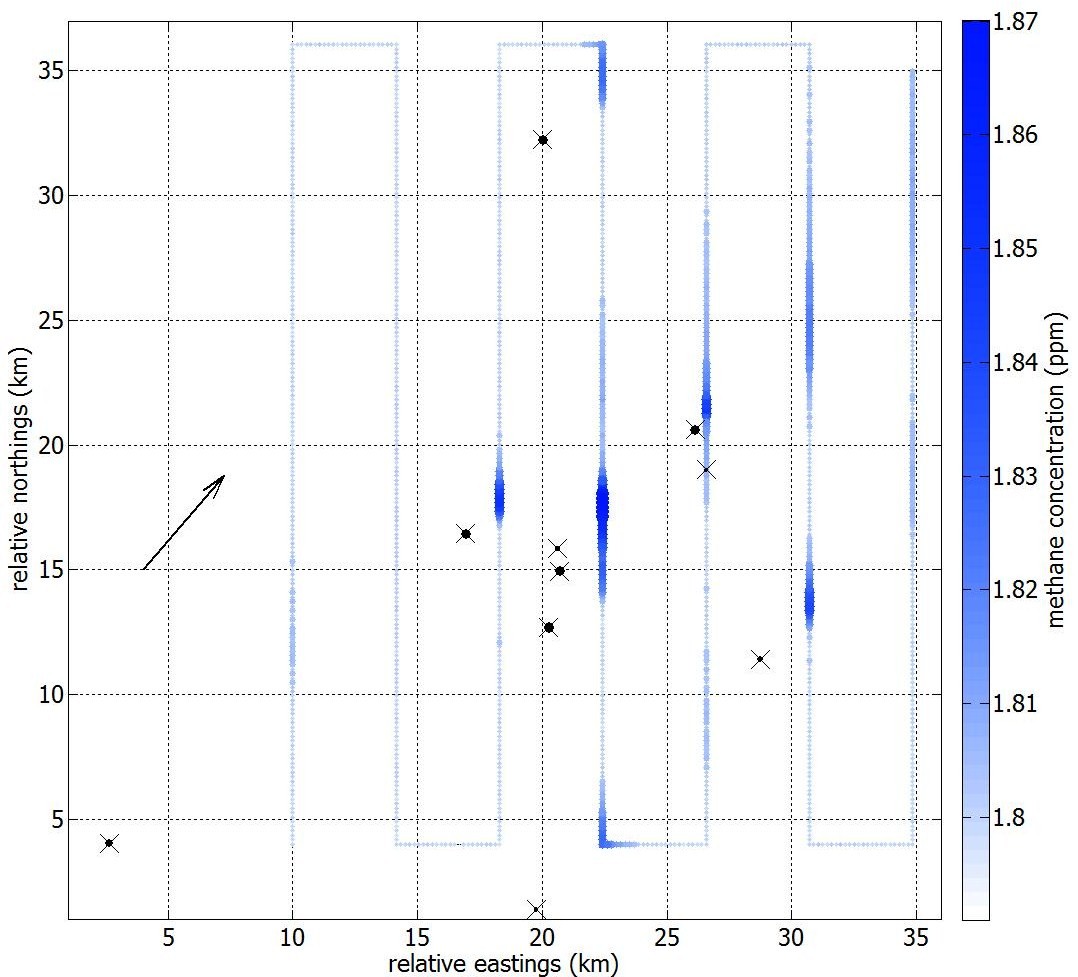}\\
  \caption{Flightpath for the synthetic problem. Marker size and colour saturation represent strength and location of simulated methane concentrations. Arrow shows direction of air movement during flight. Locations and physical extent of defined sources (each with emission rate of 0.1 $m^3s^{-1}$) are shown as crosses with overlaid black circles whose radii represent physical extent of the sources.}
  \label{Snt_SrcMapCnC}
\end{figure}

\begin{figure}[ht!]
  \center\includegraphics[width=\textwidth]{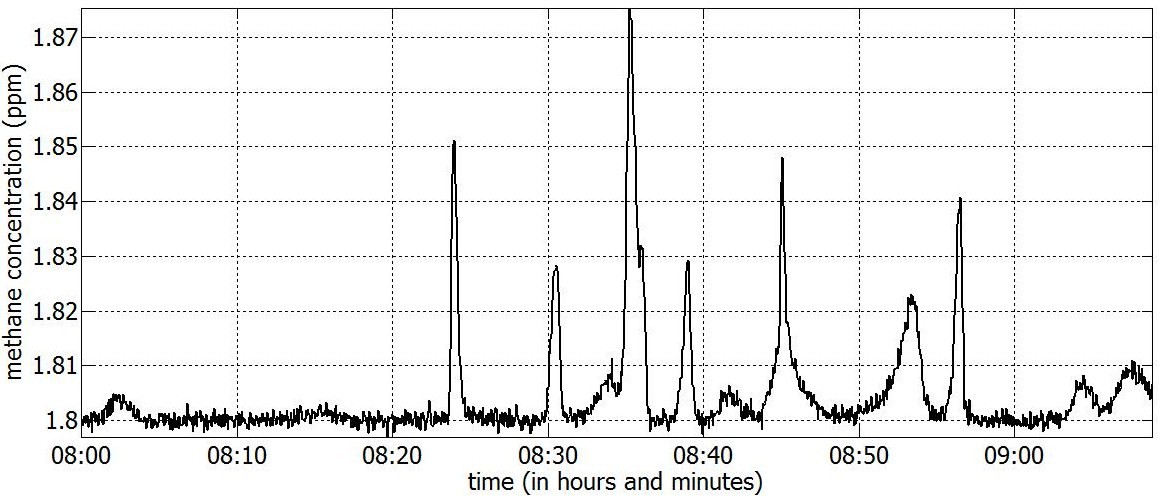}\\
  \caption{Simulated methane concentrations along flight path as a function of time for the synthetic problem.}
  \label{Snt_CnC}
\end{figure}

\begin{figure}[ht!]
  \center\includegraphics[width=\textwidth]{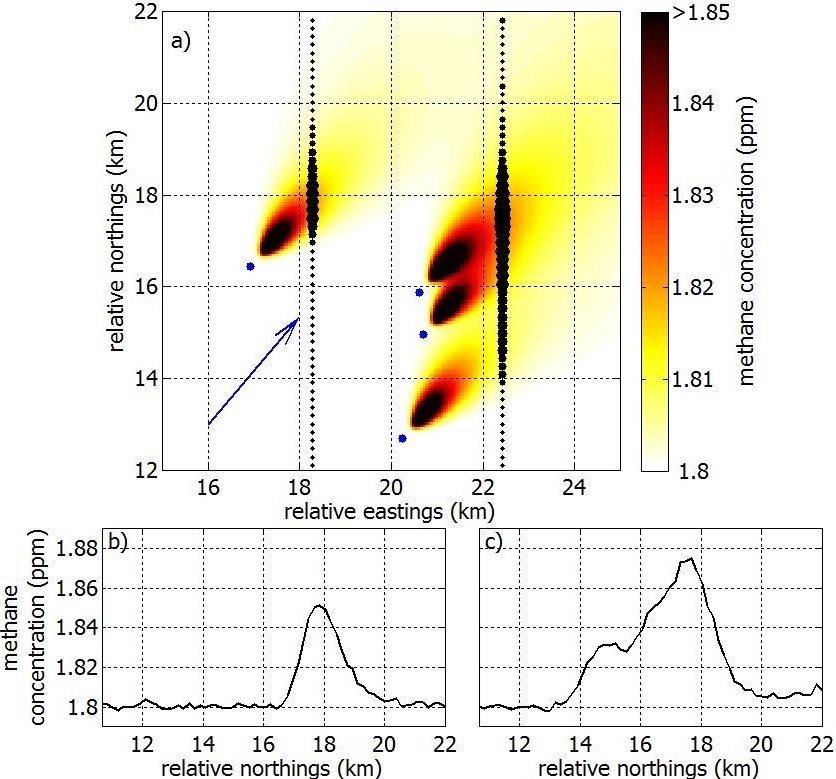}\\
  \caption{Plume concentrations for the synthetic problem: (a) Detail of central region of figure \ref{Snt_SrcMapCnC} showing simulated concentrations along flightpath as black dots, size proportional to concentration. Blue dots are also used to indicate source locations. Background concentration is $1.8$ppm. Mean direction of air movement shown by the blue arrow. (b) Simulated concentration as a function of relative northing for left--hand pass. (c) Simulated concentration as a function of relative northing for right--hand pass.}
  \label{Snt_Plm}
\end{figure}

We illustrate the approach using a synthetic problem. Gas plumes from $10$ randomly located ground level methane sources are generated with emission rate of $0.1\text{m}^3\text{s}^{-1}$ propagating in a variable wind field generated using a random walk, with wind speed and direction of $[6.3, 6.6]\text{ms}^{-1}$ at between $[218,222]^o$ meteorological, and plume opening angles of between $\gamma_H=[11.5,13.9]^o$ and $\gamma_V=[11.5,13.9]^o$, within an area of $40\text{km} \times 40\text{km}$. We simulate methane concentrations on a flight path of $1$hour $9$mins sampled at $3$s intervals yielding 1379 observations, and add a constant background of $1800$ppb. Methane concentrations and flight trajectory relative to source locations are illustrated in Figure~\ref{Snt_SrcMapCnC}. Methane concentrations in time are shown in Figure~\ref{Snt_CnC}. Detail of plume extent for two intervals of flight trajectory (corresponding to the central region of Figure~\ref{Snt_SrcMapCnC}) at $200$m above ground level, and simulated concentration measurements at the aircraft, are given in Figure~\ref{Snt_Plm}. Also shown are simulated concentrations as a function of relative northing for these trajectory intervals. Referring to pane (b), the longer right--hand tail of concentration with relative northing indicates that the aircraft's trajectory has a positive downwind component. Referring to pane (c), the source at approximately $(20,13)\text{km}$ contributes a concentration peak on the trajectory at relative northing of approximately $15\text{km}$, and the source at approximately $(20.5,16)\text{km}$ a trajectory concentration peak at approximately $17.5\text{km}$. The source at approximately $(20.5,15)\text{km}$ contributes a ``shoulder'' at approximately $16.5\text{km}$.

Inspection of Figures \ref{Snt_SrcMapCnC} and \ref{Snt_CnC} shows that each source is upwind of at least part of the flight path, and that gas emanating from each source contributes to the concentrations. Each source is therefore identifiable in principle. Some sources (e.g. those near $(20,15)\text{km}$) are close to a downwind section of the flight path, others (e.g. the source at $(3,4)\text{km}$) are relatively distant. We anticipate that the former will be more precisely modelled. Plumes from sources around $(20,15)\text{km}$ contribute to simulated concentrations on multiple downwind passes of the flight path, providing range information to resolve source location, whereas plumes from other sources (e.g. that near $(20,32)\text{km}$) intersect the flight path just once.

The survey area is partitioned into a $\by{80}{80}$ grid of ${500}$m $\times$  ${500}$m cells and a random field background model is assumed. Cell size is determined by a balance of physical (e.g. likely source extent) and computational considerations. The starting set of source locations and emission rates for RJMCMC is selected by sampling $15$ locations from the initial gridded optimisation solution and weighting each cell by its estimated emission rate. RJMCMC is executed for $13000$ iterations of which the first $3000$ burn-in iterations are ignored, generating a dependent sample from the joint posterior distribution of parameters for source locations, half-widths, emission rates, background, measurement error and wind direction bias. Burn-in length was determined by inspection of trace plots and $10,000$ iterations post burn-in were judged to be sufficient to characterise the posterior. More formal procedures (e.g. Gelman-Rubin convergence diagnostic \cite{Gelman92}) were not considered necessary.

\begin{figure}[ht!]
  %\center\includegraphics[width=\textwidth]{Snt_StrMdn.jpg}\\
  %\center\includegraphics[width=\textwidth]{Snt_LoHi.jpg}\\
  \center\includegraphics[width=0.91\textwidth]{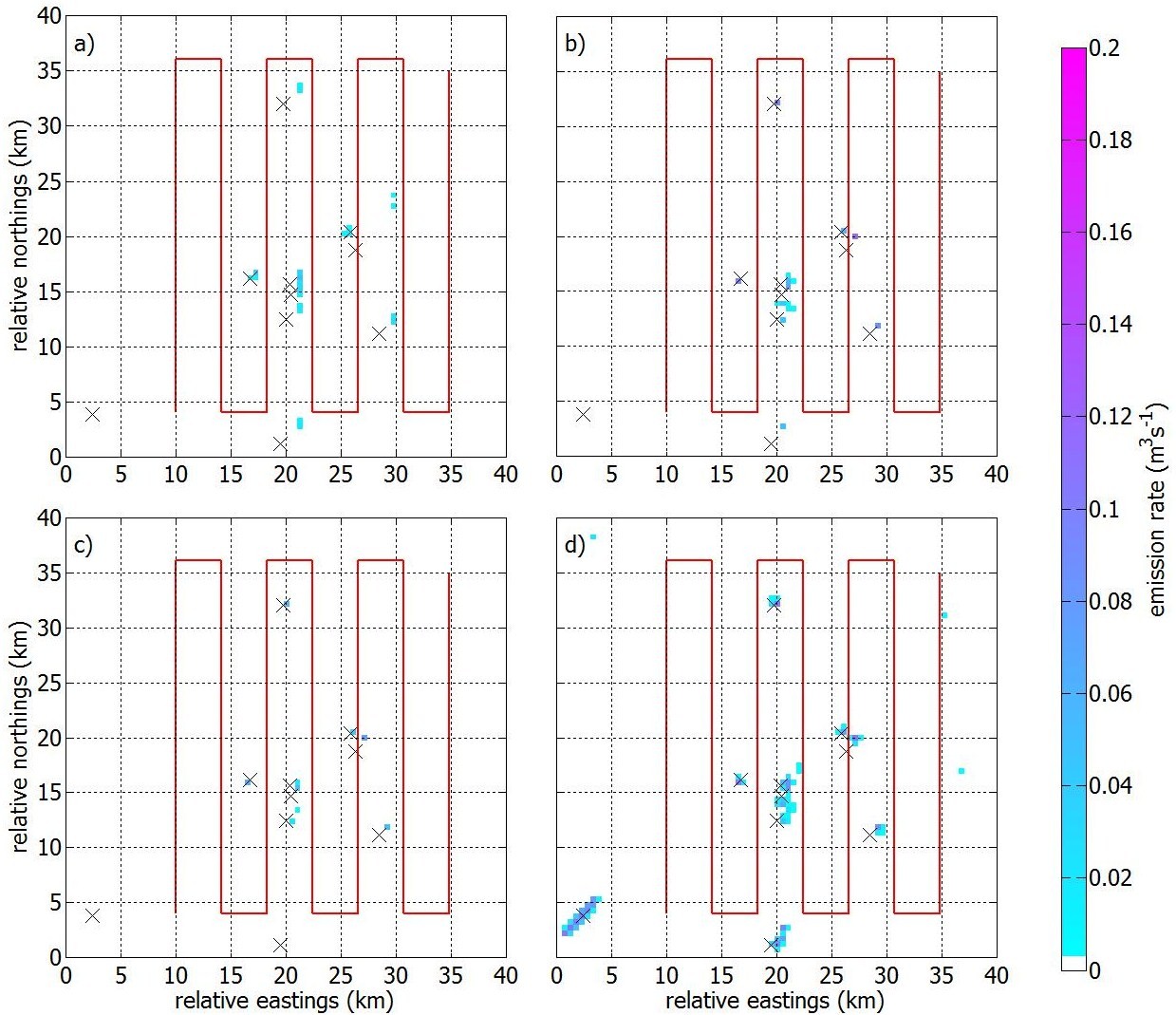}\\
  \caption{Source emission rate maps for the synthetic problem: (a) Estimate from initial optimisation, (b) Median estimate from mixture model, (c) Marginal $2.5\%$ credible value from mixture model, and (d) Marginal $97.5\%$ credible value from mixture model. For ease of comparison the mixture model results are presented on the same grid cell size as the optimsation solution. The locations of defined sources (each with emission rate of $0.1 m^3s^{-1}$) are shown as black crosses. The locations and emission rates of estimated sources are colour--coded according to the scale. The synthetic flight path is shown as a red line. }
  \label{Snt_SMLH}
\end{figure}

Estimated source emission rates are summarised in emission rate maps in Figure~\ref{Snt_SMLH}. Panel a) shows the initial optimisation solution, which identifies 9 of the 10 sources. Most estimated source locations are displaced downwind of the defined locations, closer to the flight path. The source at $(3,4)\text{km}$, furthest from the flight path, was not identified. Panel b) shows the posterior median estimate. The MCMC median solution, in common with the optimisation solution, only identifies 9 of 10 sources, though these are closer to their true locations. We summarise marginal spatial uncertainty in terms of the $2.5\%$ and $97.5\%$ credible values for source emission rates shown in Panels c) and d) respectively. 8 of the 10 known sources appear in Panel c). In Panel d) there are 3 spurious sources. For visualisation purposes only, source maps in Figure~\ref{Snt_SMLH} a) , b) and c) are estimated from gridded source estimates generated at the end of each complete iteration (over all parameters) of the Markov chain.

\begin{figure}[ht!]
  \center\includegraphics[width=0.98\textwidth]{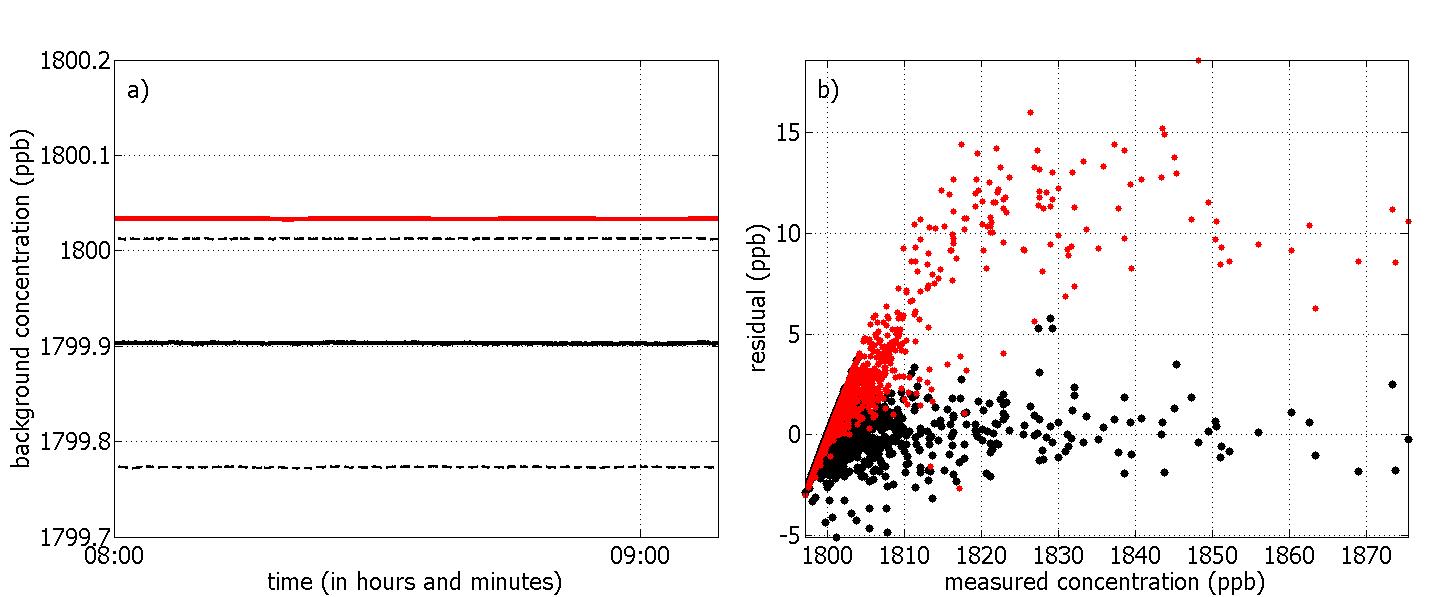}\\
  \caption{Diagnostics for the synthetic problem: (a) Estimated background concentrations along the flight path as a function of time. The initial optimisation is shown in red and the mixture model is shown in black; median (solid), $2.5\%$ and $97.5\%$ credible values (both dashed) shown. (b) Residual versus simulated methane concentrations from initial optimisation (red) and median from mixture model (black).}
  \label{Snt_BckRsd}
\end{figure}

Figure~\ref{Snt_BckRsd} (a) compares background estimates from optimisation and MCMC (with credible intervals). Recalling that the actual background is constant at 1800ppb, both estimates are within $0.1$ppb of the truth. Figure~\ref{Snt_BckRsd} (b) compares unexplained residual concentration with simulated concentration from optimisation (red) and MCMC (black). For a good model fit, we expect residuals to be zero--mean and show no relationship to the measured concentration. The MCMC residuals are relatively well distributed around zero. For optimisation, residuals corresponding to simulated concentrations close to true background (1800ppb) are small; for larger simulated concentrations, residuals are large and positive since the Laplace prior over source emission rate \eqref{E:PrpS} penalises source strength, generally resulting in positive residuals. Source locations are constrained to the centres of grid cells for the optimisation solution, but not for the mixture model estimate. The interested reader should note that the case presented is a typical example from a number of simulated cases considered but omitted for brevity.

\subsection{Landfills} \label{App:LndFll}

The analogous analysis procedure is adopted for modelling landfill measurements. For initial optimisation, the survey area is partitioned into a $\by{100}{100}$ grid of ${400}$m $\times$ ${400}$m cells. The starting point for RJMCMC is chosen by sampling $5$ locations from the optimisation solution, weighting each cell by its estimated emission rate.

\begin{figure}[ht!]
%  \center\includegraphics[width=\textwidth]{LndFll_StrMdn.jpg}\\
%  \center\includegraphics[width=\textwidth]{LndFll_LoHi.jpg}\\
  \center\includegraphics[width=0.90\textwidth]{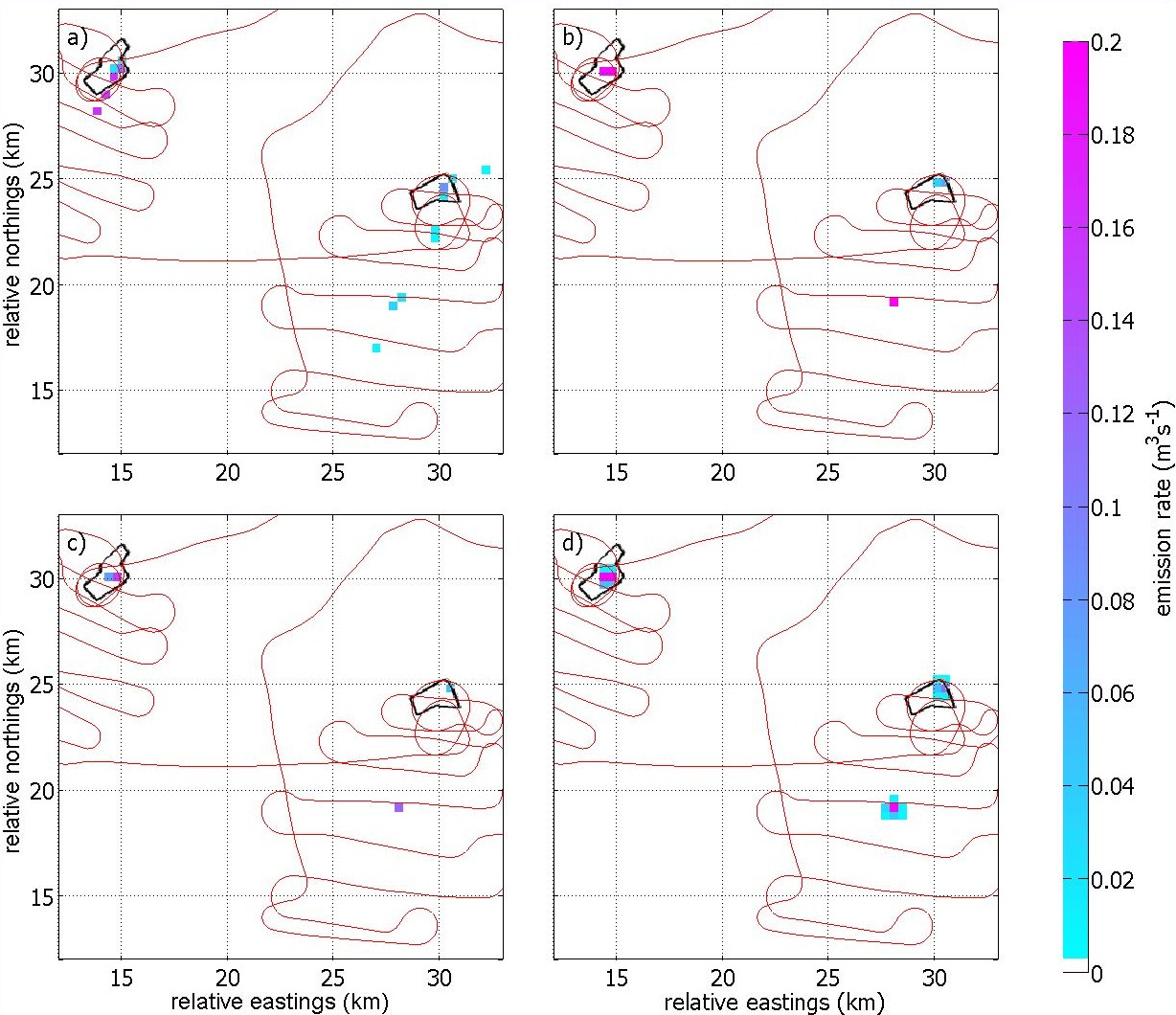}\\
  \caption{The estimated source emission rate maps for the landfill application: (a) Estimate from initial optimisation, (b) Median estimate from the mixture model, (c) Marginal $2.5\%$ credible value from the mixture model, and (d) Marginal $97.5\%$ credible value from the mixture model. All dimensions in $km$. Emission rates in $m^3s^{-1}$.  Each panel shows a common subregion of the original $40km \times 40km$ domain (referenced with respect to the origin) within which all sources are estimated. For ease of comparison the mixture model results are presented on the same grid cell size as the optimsation solution. Polygons indicate the perimeters of the landfills. The flight path is shown as a red line.}
  \label{LndFll_SMLH}
\end{figure}

\begin{figure}[ht!]
  \center\includegraphics[width=0.96\textwidth]{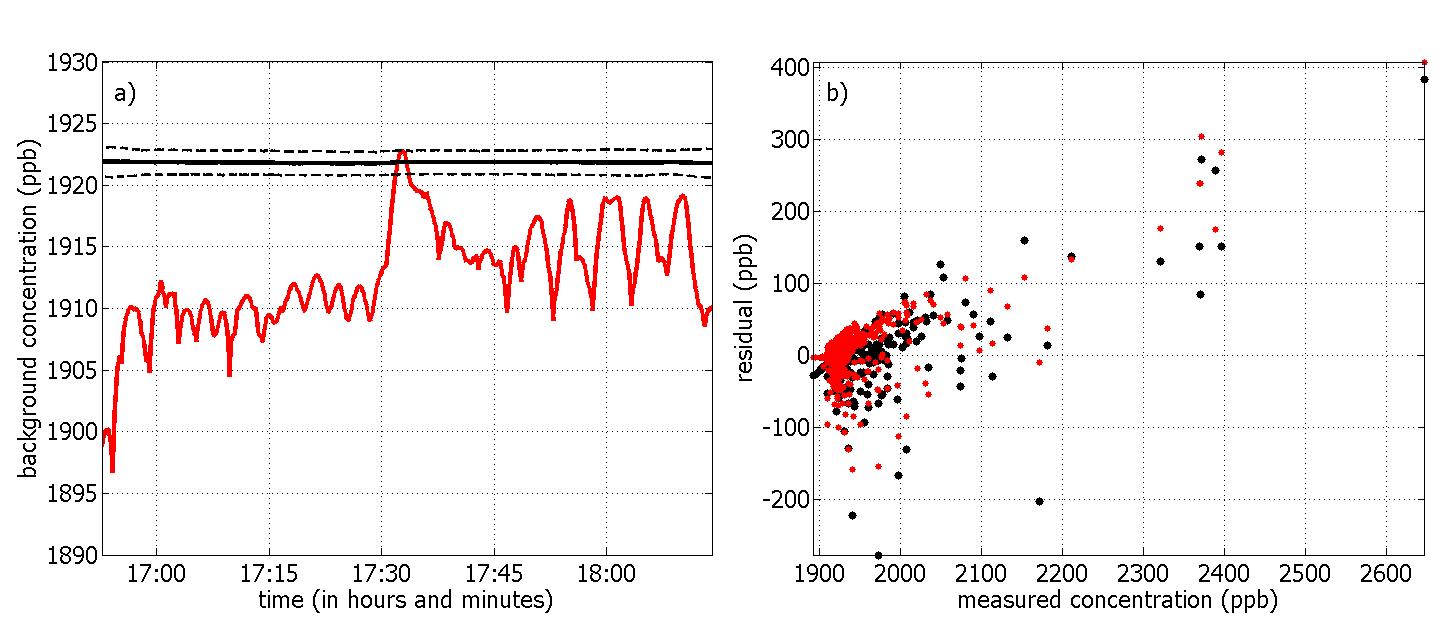}\\
  \caption{Diagnostics for the landfill application: (a) Estimated background concentrations along the flight path as a function of time. The initial optimisation is shown in red and the mixture model result is shown in black; median (solid), $2.5\%$ and $97.5\%$ credible values (both dashed) shown. (b) Residual versus measured methane concentrations from the initial optimisation (red) and median from the mixture model (black).}
  \label{LndFll_BckRsd}
\end{figure}

The estimated source emission rates are shown in Figure~\ref{LndFll_SMLH}. Panel (a) shows the initial optimisation solution; sources within the landfill boundaries are supplemented by ``ghost'' sources downwind of the landfills. These are likely due to errors in the wind field data or inadequacies in the plume model as well as changes in the wind direction during the prolonged gas transit times to the measurement locations which are up to 15km away which cannot be incorporated within the plume representation. Panel (b) shows the posterior MCMC median source estimate. Sources are centred within each landfill with just a single additional ``ghost'' source downwind of the eastern landfill $(28,19)\text{km}$, implying that correcting for wind direction bias (of approximately $2^o$) improves inversion. Panels (c) and (d) show similar characteristic to Panel (b). Interestingly, no other spurious sources appear, suggesting strong spatial localisation of sources in this case. To our knowledge there is no actual gas emission at the ``ghost'' source location in Panel (2) (which also appears in Panel (1)). If indeed the ``ghost'' source is an artefact, more sophisticated wind field corrections would be necessary to eradicate it. However, land fills are uncharacteristically strong emission sources that are detectable at extreme ranges with correspondingly extended gas transit times. For more representative sources transit times are much shorter. Figure~\ref{LndFll_BckRsd} (a) shows estimated background concentration along the flight path for initial optimisation and MCMC. Figure~\ref{LndFll_BckRsd} (b) gives fitted residuals against measured concentrations for initial optimisation and MCMC. Residuals for higher concentrations are generally larger and positive, indicating underestimation of source emission rate.

\subsection{Flare stack} \label{App:FlrStc}

\begin{figure}[ht!]
  %\center\includegraphics[width=\textwidth]{FlrStc_StrMdn.jpg}
  %\center\includegraphics[width=\textwidth]{FlrStc_LoHi.jpg}\\
  \center\includegraphics[width=0.87\textwidth]{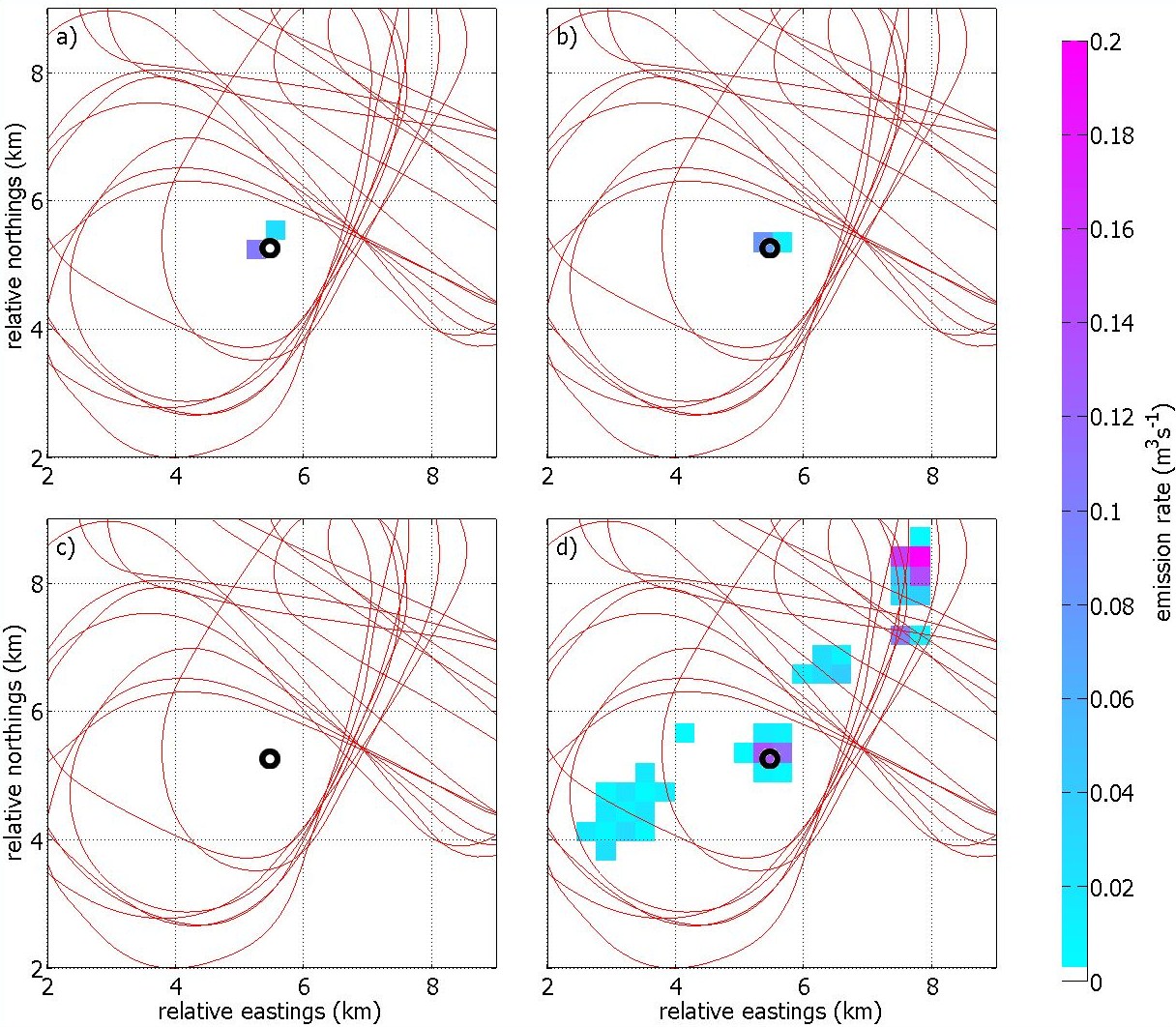}\\
  \caption{Estimated source emission rate maps for the flare stack application: (a) Estimate from the initial optimisation, (b) Median estimate from the mixture model, (c) Marginal $2.5\%$ credible value from the mixture model, and (d) Marginal $97.5\%$ credible value from the mixture model. All dimensions in $km$. Emission rates in $m^3s^{-1}$. Each panel shows a common subregion of the original $15km \times 15km$ domain (referenced with respect to the origin) within which all sources are estimated. For ease of comparison the mixture model results are presented on the same grid cell size as the optimsation solution. The black annulus indicates the location of the flare stack, a point source. The flight path is shown as a red line.}
  \label{FlrStc_SMLH}
\end{figure}

\begin{figure}[ht!]
  \center\includegraphics[width=0.94\textwidth]{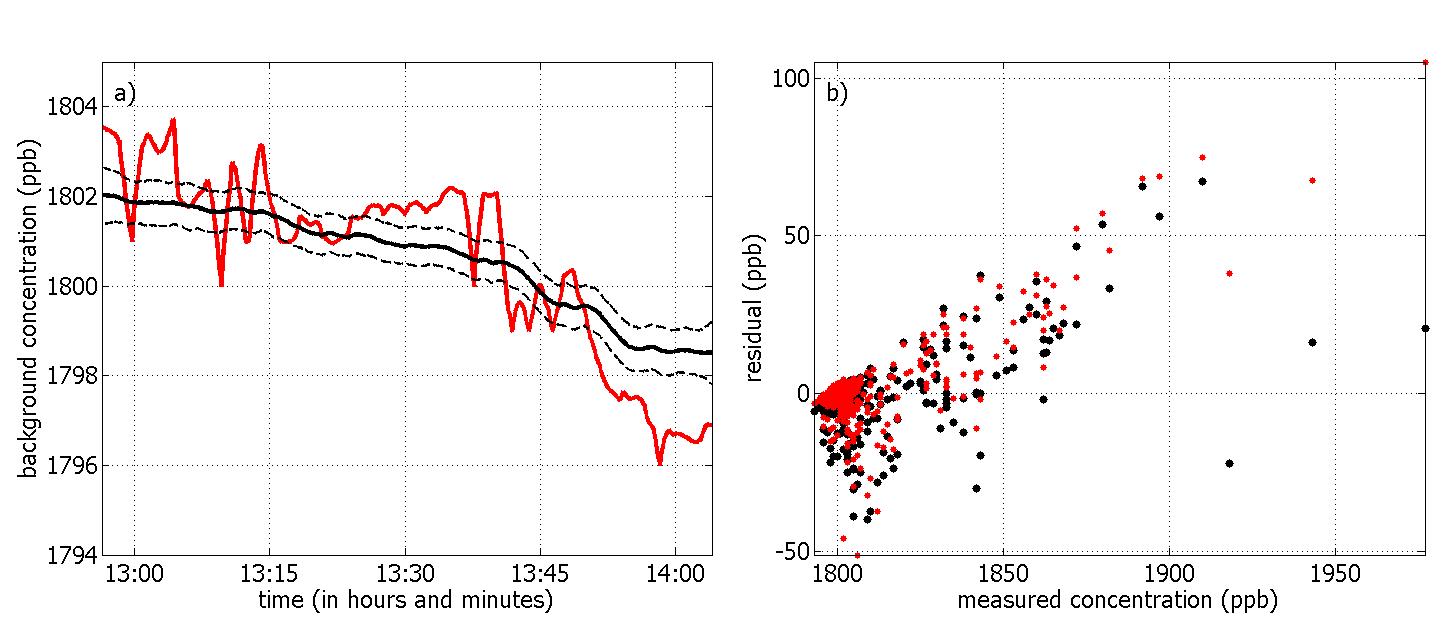}\\
  \caption{Diagnostics for the flare stack application: (a) Estimated background concentrations along the flight path as a function of time. The initial optimisation is shown in red and the mixture model result is shown in black; median (solid), $2.5\%$ and $97.5\%$ credible values (both dashed) shown. (b) Residuals versus measured methane concentrations from the initial optimisation (red) and median from the mixture model (black).}
  \label{FlrStc_BckRsd}
\end{figure}

The analysis procedure is similar to that adopted for the applications above. The survey area is partitioned into an $\by{50}{50}$ grid of ${300}$m $\times$ ${300}$m cells. The RJMCMC starting point is chosen by sampling $5$ locations from the initial optimisation solution, again weighted by emission rate.   Figure~\ref{FlrStc_SrcMapCnC} shows a clear discrepancy between plume direction and mean wind direction predicted by UKMO wind field data. The mixture model, incorporating a constant wind direction bias parameter, successfully corrects this. Inspection of Figure~\ref{FlrStc_SrcMapCnC} suggests a prior wind direction bias of approximately $-18^o$. The corresponding posterior 95\% credible interval is estimated to be $[-18.12,-17.2]^o$. This uncharacteristically large wind bias is attributed to the flight being in the late afternoon (as the ABL subsides) and is situated at the coast where winds are inherently less predictable.

Source emission rates in Figure~\ref{FlrStc_SMLH} are estimated using corrected wind directions, otherwise the initial optimisation solution (Panel (a)) would be severely compromised. The posterior median mixture model result (Panel (b)) is very similar. Panel (c) shows that 2.5\% credible values from the mixture model are $\leq0.004m^3s^{-1}$. Marginal 97.5\% credible values in Panel (d) suggest greater uncertainty in flare stack location (e.g. compared with the landfill), despite good initial optimisation and posterior median estimates. Findings from Figure~\ref{FlrStc_BckRsd} are similar to those from Figure~\ref{LndFll_BckRsd}.
%%%%%%%%%%%%%%%%%%%%%%%%%%%%%%%%%%%%%%%%%%%%%%%%%%%%%%%%%%%%

%%%%%%%%%%%%%%%%%%%%%%%%%%%%%%%%%%%%%%%%%%%%%%%%%%%%%%%%%%%%
%Discussion and conclusions
\section{Discussion} \label{Dsc}
Detection and location of sources of gas emissions into the atmosphere is a topic of intense interest. In this work we describe a method for detecting, locating and quantifying such sources using remotely obtained gas concentration data. The method developed is broadly applicable to any gas or passively transported, detectable atmospheric component, such as aerosols, radon, smoke, ash, dust and viruses. The method can be applied using ground based or airborne concentration data collected using point or line--of--sight sensors. Here, atmospheric point concentration measurements are modelled as the sum of a spatially and temporally smooth atmospheric background, augmented by concentrations arising from local sources. We model source emission rates with a Gaussian mixture model and use a Markov random field to represent the atmospheric background concentration component of the measurements. A Gaussian plume atmospheric eddy dispersion model represents the gas dispersion process between sources and measurement locations. An initial point estimate of background concentration and source emission rates is obtained using mixed $\ell_2-\ell_1$ optimisation over a discretised grid of potential source locations. Subsequent reversible jump Markov chain Monte Carlo inference provides estimated values and uncertainties for the number, emission rates, locations and areas of sources, as well as atmospheric background concentrations and other model parameters. We investigate the performance of the approach first for a synthetic inversion problem. We then apply the model to locating and detecting sources of gas emissions using actual airborne data (a) in the vicinity of two landfills, and (b) in the vicinity of a gas flare stack. All analysis was performed using \cite{MATLAB2011}.

As discussed in the introduction, individual model components and inference tools used in this work are commonplace in the modelling literature. The combination of components and tools, pulling together physical constraints with rigourous analysis, has proved useful for the remote sensing applications considered in this work. Nevertheless, to the best of our knowledge, this is one of the first applications of Bayesian inference using reversible-jump MCMC to simultaneous multiple source and smooth spatio--temporal background estimation. The Gaussian plume model is a particularly simple steady--state approximation to dispersion of a gas release into the atmosphere, widely used in the environmental modelling literature (e.g. \cite{Gifford76}). In this work, the plume model provides a reasonable basis for estimating known flare stack and landfill locations. However, we find that correcting bias in (predicted) wind directions supplied by UKMO improves inference.  There is evidence, in the form of a ``ghost'' source downwind of the eastern landfill, that a simple bias correction is not adequate, and that a more sophisticated approach (e.g. a slowly varying wind-direction bias) might be beneficial. There is considerable opportunity to achieve this within the Bayesian modelling framework. Incorporating plume model uncertainty in the initial optimisation can be achieved in some sense by considering optimisation over a representative set of forward model matrices $A$ (in \eqref{E:MdlY}), rather than a single choice.  Predicated on the availability of wind field data of adequate quality we might also consider more sophisticated plume models, e.g. plumes following wind flow lines, or from computational fluid dynamics. In the MCMC case, we assume a-priori that sources can occur with equal probability at any location. For RJMCMC we sample those grid locations from the optimisation solution with the greatest emission rates as a starting solution. Work continues to explore incorporation of spatial prior distributions for source location, for example using Polya trees to encode some degree of source clustering. There is scope to develop more sophisticated background models incorporating parameters known to influence background methane concentration (such as topography). It would be interesting to explore modelling background as the superposition of a number of distant sources. Our field experience suggests that natural gas seeps can be intermittent, requiring adaptation of our model formulation. Smoothly varying gas release rates could be accommodated relatively simply.

The Gaussian plume model provides an elementary means of modelling gas transport from source to measurement location under ideal steady state wind field assumptions, allowing rapid estimation of forward model matrices $A$ at the expense of accuracy and precision. Given inherent uncertainties in the estimates of wind field parameter values supplied by UKMO, we consider the Gaussian plume adequate for the purposes of the current work. For example in the landfills application, assuming ideal wind field conditions over an interval of approximately $10$ minutes for wind speed of approximately $6.5\text{ms}^{-1}$, suggests that the plume model is appropriate for measurement locations within approximately $4\text{km}$ downwind of a source. More distant measurement locations require ideal conditions over longer periods for the Gaussian plume model to be appropriate. Nevertheless, even over larger distances, the Gaussian plume is likely to provide reasonable approximation to reality provided the wind field remains relatively steady.

As implemented in the current work, optimisation is used to provide an initial point solution for inversion on a spatial grid. Subsequent Bayesian inference gives a more flexible grid-free mixture model framework within which to estimate the joint posterior distribution of all parameters, providing in particular estimates for parameter uncertainty. Early attempts at inversion followed a stepwise approach in which atmospheric background was estimated prior to, and independent of emission sources. The current approach, involving simultaneous estimation of background, sources and wind field characteristics improves performance. We also explored Bayesian inference on the same spatial grid used for the initial optimisation. The very large number of potential source locations makes this computationally intensive.

Our experience of processing multiple survey datasets has made clear the need for rigorous data management and pre-processing procedures, e.g. in the merging of spatio-temporal data from different sources (e.g. aircraft and UKMO). Efficiency of inference can be improved for a given deployment by specifying a flight trajectory (or sequence of flight trajectories) appropriately, given prevailing wind conditions and prior information concerning likely source locations. Methods of statistical experimental design are central to achieving this for both airborne and ground based line--of--sight gas sensors. It is also strongly desirable to have a means of confirming the quality of inference, particularly of source location and release rate, using a persistent known gas source within the region of interest. In some cases, this might take the form of an existing methane source (such as a flare stack or landfill), or perhaps a small controlled release (e.g from a gas cylinder).
%%%%%%%%%%%%%%%%%%%%%%%%%%%%%%%%%%%%%%%%%%%%%%%%%%%%%%%%%%%%

%%%%%%%%%%%%%%%%%%%%%%%%%%%%%%%%%%%%%%%%%%%%%%%%%%%%%%%%%%%%
%Acknowledgement
\section{Acknowledgement} \label{Ack}
The authors gratefully acknowledge the long--term co--operation of Sander Geophysics Limited, Ottawa, Canada and discussions with colleagues at Shell and at MIT's Laboratory for Information and Decision Systems. \\
%%%%%%%%%%%%%%%%%%%%%%%%%%%%%%%%%%%%%%%%%%%%%%%%%%%%%%%%%%%%

\pagebreak

%%%%%%%%%%%%%%%%%%%%%%%%%%%%%%%%%%%%%%%%%%%%%%%%%%%%%%%%%%%%
%Appendix
\begin{appendices}

\section{Background model} \label{App:Bck}
\subsection{MRF background model}\label{ApxMRFBack}

We model the background as a Gauss-Markov random field, a class of graphical model. Markov random fields are joint distributions for variables $X_1,\ldots,X_N$ and an associated undirected graph $G=(\mathscr{E},\mathscr{V})$. The vertex set $\mathscr{V}:=\{1,\ldots,N\}$, and the edge set $\mathscr{E}$ is a subset of $\mathscr{V}\times\mathscr{V}$. The graph specifies the conditional independence structure of random variables as follows. If three sets of variables $A,B,C\subseteq\mathscr{V}$ are such that the set $B$ separates $A$ from $C$ in the graph $G$, then $(X_i)_{i\in A}$ must be conditionally independent of $(X_i)_{i\in C}$ given $(X_i)_{i\in B}$. A simple special case of a Markov random field is a Markov chain, wherein $G$ is simply a linear graph. In a Gauss-Markov random field, the random variables are also assumed to be jointly Gaussian. It can be shown (e.g. \cite{Speed1986}) that for Gaussian variables, the conditional independence condition is equivalent to the precision matrix $J$ being sparse with respect to the graph $G$. That is, for $i$ different from $j$, $J_{ij}\ne 0$ if and only if there is an edge between node $i$ and node $j$ in $G$.

For the random field model, we particularise \eqref{E:PrpBta} as:
\begin{equation} \label{E:PrpBta2}
f(\bta) \propto \exp\{ - \tfrac{1}{2} \bta^T \J_{\bta} \bta \}
\end{equation}
where again $\J_{\bta}$ is sparse with respect to a graph to be designed. As stated in Sec.~\ref{Mdl:BckMdl}, we take $\P$ to be the identity matrix, so that our background estimate $\myb$ is simply equal to $\bta$.

\begin{figure}[ht!]
  \center\includegraphics[width=1\textwidth]{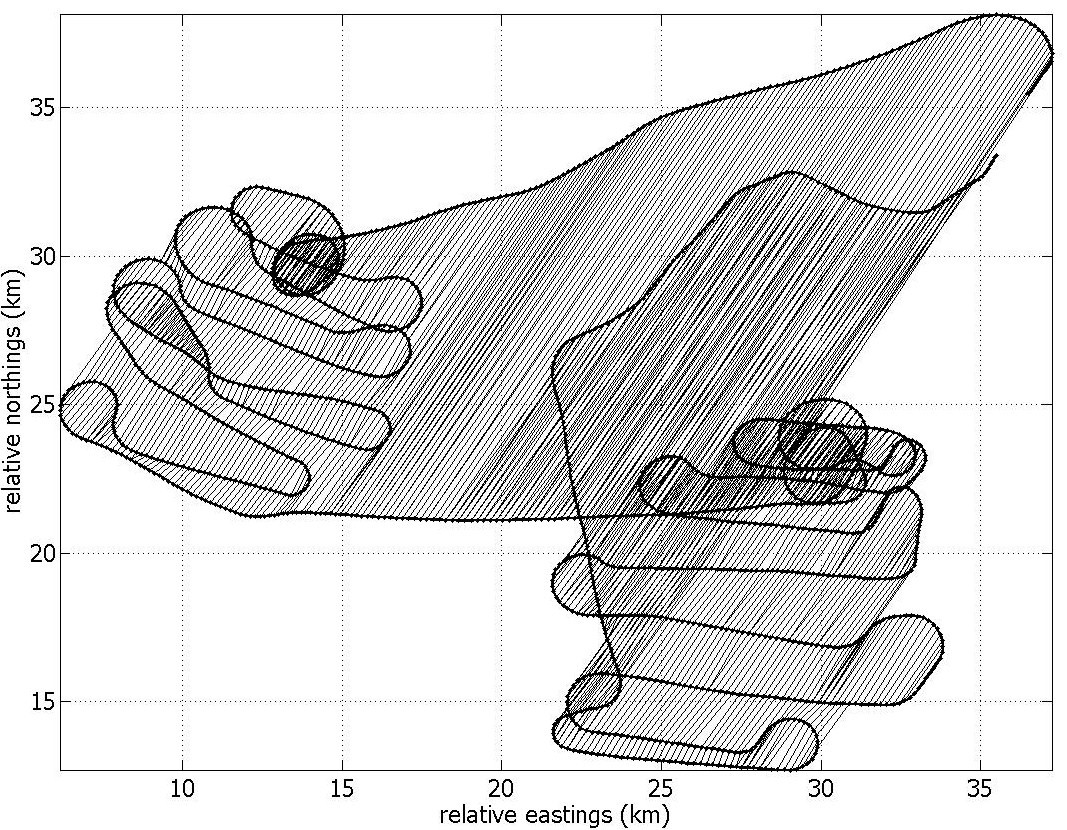}\\
  \caption{Illustration of ``wind--linked'' points along the flight trajectory for the landfill application}
  \label{LndFll_WndLnks}
\end{figure}

In designing a Gauss--Markov random field to model the background field, we seek to capture two effects. First, the background should be smooth. Second, since the background concentration travels with wind, it should be smoother along the direction of the wind. To model these two effects, we introduce two different types of edges in the graphical model. The first are edges connecting adjacent measurements vertices. These ensure overall smoothness of the background with respect to time and space. The second type of edge concerns the wind. At each measurement point, we consider the line along the wind direction from this point. We find the next measurement point that crosses this line, and connect an edge in the graph between these two points. Thus the second point is as close as possible to directly in line with the wind from the initial point. ``Wind--linked'' points along the trajectory for the landfill application is given in Figure~\ref{LndFll_WndLnks}.

The graphical structure outlined above determines the sparsity pattern of $J_{\bta}$, but not the values themselves. The structure of $J_{\bta}$ is as follows:
\[J_{\bta}=\sum_{(i,j)\in\mathscr{E}}\alpha_{ij}\Lambda_{ij}\]
where $\Lambda_{ij}$ is a matrix non-zero only at the intersections of the $i$th and $j$th rows and columns, and the non-zero elements of $\Lambda_{ij}$ are given by:
\[\left[\begin{array}{cc}1&-1\\-1&1\end{array}\right].\]
This may be alternately stated as, for any vector $x$:
\[x^T J_{\bta} x = \sum_{(i,j)\in\mathscr{E}}\alpha_{i,j}(x_i-x_j)^2.\]

The values $\alpha_{ij}$ represent the strengths of the links, and they are chosen as follows. We assume that if the background value were measured in the same packet of air at two different times, the change in background concentration would roughly grow with the time difference. Similarly, at two different points in space at the same time, the difference in background concentration would grow with the spatial separation. This suggests that the strength of each link should be determined by the time difference and the spatial distance between the two connected points. In particular, we take as parameters two constants $c_T$ and $c_D$. The parameter $c_T$ is the expected change in background concentration per second difference in measurement time, and $c_D$ is the change in background concentration per meter separation. The strength $\alpha_{ij}$ is then given by:
\[\alpha_{ij}=\frac{1}{(c_T\,\Delta T+c_D\,\Delta D)^2}\]
where $\Delta T$ is the difference in time between the two measurements and $\Delta D$ is the spatial distance between the two measurements. For the links between adjacent measurements, we calculate $\Delta D$ as simply the distance from first measurement point to the second measurement point. For the wind-based cross-links, we use the distance from the second measurement to the point in space where the packet of air at the first measurement would be, assuming that it travels with the wind at the wind speed and direction as measured at the first point. Other parametric forms for $\alpha_{i,j}$, e.g. motivated by diffusion--advection transport (see \ref{Apx:PolyBack} below), could be considered.

\subsection{Polynomial background model}\label{Apx:PolyBack}

Based on the idea that the background field is slowly varying in space and time, we can also construct a basis for a smooth, low-dimensional manifold using low-degree Chebyshev polynomials.  If desired, it can be trivially extended to more dimensions to account for  regression on other variables, such as pressure, temperature, etc.

The background field is modelled as a function of space $x,y,z$ and time $t$, and coefficient vector $\bta$:
\begin{equation*}\label{polyback}
b(x,y,z,t; \, \bta) = \sum_{i,j,k,l} \ \beta_{i,j,k,l} T_i(x)T_j(y)T_k(z)T_l(t)
\end{equation*}
where $T_n$ is the Chebyshev polynomial of the second kind of order $n$, with the coordinates properly transformed such that each dimension $x,y,z$, and $t$ are mapped onto the interval $[-1,1]$.

The unscaled Chebyshev polynomials of the second kind \cite{Cheby85} are defined by the recurrence relation ($x\in[-1,1]$)
\begin{align*}
 U_0(x) &= 1, &U_1(x) &= 2x,& U_{n+1}(x) &= 2xU_n(x) - U_{n-1}(x) \,.
\end{align*}

One of the advantages of the Chebyshev polynomial model is that it has an analytical form, and therefore it is very easy to evaluate the function and its derivatives on any point in space, particularly those outside the flight track.

The regularisation and conditioning of this type of background consists of three components:
\begin{description}

\item[Regularisation with respect to a reference background.] A constant reference background level $b_0 = P\bta_0$ can be used for regularisation. Penalisation for deviation of $b$ from $b_0$ can be written as the  quadratic form
\[
 R_1(\bta) = (\bta - \bta_0)^T I (\bta - \bta_0) \,.
\]

\item[Smoothness regularisation.] We choose as a smoothness measure the sum of squares of the curvature of the background field evaluated at $\mathcal{G} = \{\xi_i = (x_i,y_i,z_i,t_i)\}_i$, a discretisation of the spatio-temporal domain.
\begin{align}
R_2(\bta) &= \sum_{\xi_i \in \mathcal{G}} \left[ \Big( \diff{^2b}{x^2} \Big)^2 + \Big( \diff{^2b}{y^2} \Big)^2 + \Big( \diff{^2b}{z^2} \Big)^2 + \Big( \diff{^2b}{t^2} \Big)^2 \right]_{\xi = \xi_i} \label{smoothnessterm}
\intertext{and the analytical form \eqref{polyback} allows us to easily write \eqref{smoothnessterm} as a quadratic form:}
  &= (\bta - \bta_0)^T J_2 (\bta - \bta_0) \nonumber
\end{align}
where $J_2$ is a symmetric semi-definite matrix.

\item[Wind transport regularisation.] The background field can be made to obey the transport equation along the wind field $\w(x,y,z,t)$:
\begin{equation*}
  \w \cdot \nabla_{x,y,z} b + \diff{b}{t} \approx 0 \label{windtransport}
\end{equation*}

Similar to \eqref{smoothnessterm}, we formulate the wind transport penalty as a quadratic form
\begin{align*}
  R_3(\bta) &= \sum_{\xi_i \in \mathcal{G}} \left[\w(\xi_i) \cdot \nabla_{x,y,z} b(\xi_i) + \diff{b}{t}(\xi_i) \right]^2 \notag \\
  & = (\bta - \bta_0)^T J_3 (\bta - \bta_0) \label{windpenalty} \,,
\end{align*}
where $J_3$ is a symmetric semi--definite matrix.
\end{description}

In the end, the three penalisations can be added as a single quadratic form
\[
(\bta - \bta_0)^T J (\bta - \bta_0) \,,\qquad \text{where } J = \mu_1 I + \mu_2 J_2 + \mu_1 J_3 \,,
\]
and $\mu_1$, $\mu_2$, and $\mu_3$ are relative weights that allow an expert to choose the strength of each regularisation term.
%%%%%%%%%%%%%%%%%%%%%%%%%%%%%%%%%%%%%%%%%%%%%%%%%%%%%%%%%%%%

%%%%%%%%%%%%%%%%%%%%%%%%%%%%%%%%%%%%%%%%%%%%%%%%%%%%%%%%%%%%
\section{Initial parameter estimation} \label{App:IntPrmEst}
We solve the optimisation problem \eqref{E:ObjFnc} based on an alternating direction method.  The objective function is not entirely separable with respect to the variables $\s$ and $\bta$, since they are coupled through the data fitting term $\|A\s + P\bta -\y\|^2$, but the rest of the terms and the constraints are uncoupled.  These type of split methods were devised originally \mycite{Masao1992} for large-scale separable problems, but variants have gained a lot of attention recently to solve other problems, especially in the fields of image processing, machine learning and compressed sensing \mycite{Yin2008,Yang2009,Deng2011} where the objective function can be written as the sum of convex subproblems, each with special structure and characteristics that can be exploited, such as the sum of smooth and non-smooth terms (like the $\ell_2$-$\ell_1$ problem that we have).  Even if the optimisation problem can be written as a standard linear or quadratic program --and their respective theories and solvers are very mature-- these new methods can have better practical and computational properties.

In our case, the main optimisation problem \eqref{E:ObjFnc} is indeed a quadratic objective function with linear inequality constraints, and $J$ is a positive semi-definite matrix, so it could be solved with standard convex quadratic solvers.  However, the problem has a special structure that can be exploited and we have found that a iterative split technique converges much faster than using Matlab's \texttt{quadprog} routine\footnote{Matlab 2008a}.

The algorithm can be described as an alternating sequence of optimisation of the background and ground sources.  In the first stage, \eqref{E:ObjFnc} is minimised only  with respect to $\bta$, the parameterised background, while keeping the current estimate of $\s$ fixed. Analogously, in the second stage, \eqref{E:ObjFnc} is minimised only with respect to the ground sources emission rate $\s$, while keeping the current estimate of the background fixed. Then,  a criterion of optimality is computed and, if satisfied, the current solution is deemed to be close enough to the global minimum.  If it is not satisfied, the algorithm iterates again through the first and second stages.

\begin{description}
\item[Step 1: Estimate the background.]
Although it is physically reasonable, the non-negativity constraint of the background, $\vect{0} \leq P\bta$, is actually not enforced explicitly, since in practice it is never active. The sub-problem
\begin{subequations}
\begin{align}
  \min_{\bta, \vect{w}} & \qquad \tfrac{1}{2\sigma_\epsilon^2} \|A\s + P\bta -\y \|^2 + \tfrac{\mu}{2} (\bta-\bta_0)^T J (\bta - \bta_0) \nonumber \\
  \text{subject to } & \qquad P\bta \leq \y + \tau\,. \nonumber
\end{align}\label{LToptprbbeta}
\end{subequations}
is solved by the method of augmented Lagrangian for inequalities \mycite{NW1999}. To do this, the inequality constraint is converted into an equality constraint by introducing a slack non-negative variable $\vect{w}$:
\begin{align*}
  \min_{\bta, \vect{w}} & \quad \tfrac{1}{2\sigma_\epsilon^2} \|A\s + P\bta -\y \|^2 + \tfrac{\mu}{2} (\bta-\bta_0)^T J (\bta - \bta_0) \\
  \text{subject to } & \quad
  \begin{cases}   - P\bta + \y + \tau +  \vect{w} = 0  \\
      \vect{w}  \geq 0
  \end{cases}
\end{align*}

This, in turn, is solved by moving the equality constraint into the objective function and introducing its corresponding Lagrange multiplier  $\z\in\R^n$, to obtain
\begin{align*}
  \min_{\bta, \vect{w}} & \quad L(\bta,\z ; \eta) = \tfrac{1}{2\sigma_\epsilon^2} \|A\s + P\bta -\y \|^2 + \tfrac{\mu}{2} (\bta-\bta_0)^T J (\bta - \bta_0) + \sum_{j=1}^n \psi\big( c_i( \beta ) , z_i ; \eta )  \\
  \text{subject to } & \quad
      \vect{w}  \geq 0
\intertext{where $c_i$ is the $i$-th component of the constraint}
  \vect{c} &= - P\bta + \y + \tau +  \vect{w}
\intertext{and $\psi$ is the auxiliary function defined as}
\psi(a,b ; \eta) &= \begin{cases}
                     -b a + \frac{1}{2\eta}a^2 & \text{if } \ a-\eta b \leq 0 \,,\\
		     -\frac{\eta}{2} b^2 & \text{otherwise.}
                    \end{cases}
\end{align*}
and this problem is solved with a combination of Newton's method and projected gradient to maintain the feasibility $\vect{w} \geq 0$.
For more details on the practical considerations with this type of methods, see \S 17.4 of \mycite{NW1999}.

\item[Step 2: Estimate the sources.]
The subproblem
\begin{subequations}
\begin{align}
  \min_{\s} & \qquad \tfrac{1}{2\sigma_\epsilon^2} \|A\s + P\bta -\y \|^2 + \lambda\|Q\s\|_1 \nonumber \\
  \text{subject to} & \qquad 0 \leq \s \leq s_{\max} \nonumber
\end{align}\label{LToptprbs}
\end{subequations}
has simple bound constraints and we use a gradient-projection method to solve it.  We use a majorise-minimise\mycite{Majorize} method to solve for this quadratic form with simple bound constraints $0\leq \s\leq s_{\max}$.
\begin{equation*}
\min_{\s \geq 0} \quad \tfrac{1}{2\sigma_\epsilon^2} \|A\s + P\bta -\y\|^2 + \tfrac{\mu}{2} (\bta-\bta_0)^T J (\bta - \bta_0) + \lambda \vect{q}^T \s
\end{equation*}
Note that the term $\|Q\s\|_1$ has been replaced by $\vect{q}^T\s$, where $\vect{q}=\mathrm{diag}(Q)$. This is due to the fact that since we have that  $\|\s\|_1 \equiv \vect{1}^T\s$ when $\s \geq 0$, and therefore the term is fully differentiable in the half-space $\s \geq 0$, and no concerns of the non-differentiability of the $\ell_1$-norm must be taken into account.

\end{description}
%%%%%%%%%%%%%%%%%%%%%%%%%%%%%%%%%%%%%%%%%%%%%%%%%%%%%%%%%%%%

%%%%%%%%%%%%%%%%%%%%%%%%%%%%%%%%%%%%%%%%%%%%%%%%%%%%%%%%%%%%
\section{Mixture model} \label{App:MxtMdl}
\subsection{MCMC step types}

Referring to section \ref{Mdl:MxtMdl}, we write the parameter set as $\tht$ for brevity. $\tht$ consists of source parameters $\z$, $\w$, $\s$, background parameters $\bta$, measurement error standard deviation $\sigma_{\epsilon}$ and (potentially) wind direction and other plume bias and uncertainty terms.

\subsubsection*{Metropolis Hastings}

In conventional MCMC using the Metropolis Hastings algorithm, we construct a Markov chain with stationary distribution corresponding to the posterior distribution $f(\tht | \D)$ given observed data $\D$ consisting of observed airborne concentrations $\y$ (and potentially wind field measurements). For transitions between two states $\{\thtkpp,\thtnotkpp\}$ and $\{\thtkpp',\thtnotkpp\}$, we ensure reversibility of the Markov chain by imposing the detailed balance condition:

\begin{equation*} \label{E:DtlBln1}
\int f(\thtkpp | \D, \thtnotkpp) q(\thtkpp,\thtkpp' | \thtnotkpp) \alpha(\thtkpp,\thtkpp' | \thtnotkpp) d\thtkpp = \int f(\thtkpp' | \D, \thtnotkpp) q(\thtkpp',\thtkpp | \thtnotkpp) \alpha(\thtkpp',\thtkpp | \thtnotkpp) d\thtkpp'
\end{equation*}

for acceptance probability $\alpha$ and proposal distribution $q$, from which \eqref{E:MtrHst1} emerges. In the current work, the proposal distribution $q$ always corresponds to a random walk such that $q(\thtkpp, \thtkpp'|\thtnotkpp)=q(\thtkpp', \thtkpp|\thtnotkpp)$ (e.g. we might use $\thtkpp' = \thtkpp + \epsilon$ for (multivariate) Gaussian $\epsilon$). The manner in which the chain explores the posterior distribution for dependent variables or multi-modal distributions (such as our mixture model) can be improved by updating subsets of parameters in a particular order and adopting a good starting solution.

\subsubsection*{Reversible jump Metropolis Hastings}

In RJMCMC, we construct a Markov chain which satisfies an extended balance equation:

\begin{equation*} \label{E:DtlBln2}
\int f(\thtkpp | \D, \thtnotkpp) g(\ph|\thtnotkpp) j(\thtkpp|\thtnotkpp) \alpha(\thtkpp,\thtkpp'|\thtnotkpp) d\thtkpp d\ph= \int f(\thtkpp' | \D, \thtnotkpp) g'(\ph'|\thtnotkpp) j(\thtkpp'|\thtnotkpp) \alpha(\thtkpp',\thtkpp|\thtnotkpp) d\thtkpp'd\ph'
\end{equation*}

where $\thtkpp$ and $\thtkpp'$ are now of different dimensions . We facilitate dimension-jumping by constructing augmented sets of variables $\{\thtkpp,\ph\}$ and $\{\thtkpp',\ph'\}$ which are of the same dimension, using specified bijective functions $h$ to move between them, such that $(\thtkpp',\ph')=h(\thtkpp,\ph)$ (and $(\thtkpp,\ph)=h'(\thtkpp',\ph')$, with $h' \equiv h^{-1}$ ). We also specify the joint distribution $g(\ph|\thtnotkpp)$ (typically uniform or Gaussian; $g'(\ph'|\thtnotkpp)$ is known given $g$ and $h$). $j(\thtkpp|\thtnotkpp)$ is the probability that a particular dimension-jumping move will be attempted from state $\{\thtkpp,\thtnotkpp\}$.

The Metropolis-Hastings acceptance probability $\alpha(\thtkpp,\thtkpp'|\thtnotkpp)$ (corresponding to \eqref{E:MtrHst1}) for a dimension--jumping move becomes:

\begin{equation} \label{E:MtrHst2}
\alpha(\thtkpp,\thtkpp'|\thtnotkpp) = \max \left\{1,  \frac{f(\thtkpp' | \D, \thtnotkpp) j(\thtkpp'|\thtnotkpp) g'(\ph'|\thtnotkpp)}{f(\thtkpp | \D, \thtnotkpp) j(\thtkpp|\thtnotkpp) g(\ph|\thtnotkpp)} \left | \diff{(\thtkpp',\ph')}{(\thtkpp,\ph)} \right |  \right\}
\end{equation}

where the Jacobian is easily evaluated given $h$. The acceptance probability for the reverse move is similarly:

\begin{equation*} \label{E:MtrHst2r}
\alpha(\thtkpp',\thtkpp|\thtnotkpp) = \max \left\{1,  \frac{f(\thtkpp | \D,\thtnotkpp) j(\thtkpp|\thtnotkpp) g(\ph|\thtnotkpp)}{f(\tht' | \D, \thtnotkpp) j(\thtkpp'|\thtnotkpp) g'(\ph'|\thtnotkpp)} \left | \diff{(\tht,\ph)}{(\tht',\ph')} \right |  \right\}
\end{equation*}

We use RJMCMC to update the number of sources $m$ by independent birth-death and coalesce-split steps.

\subsubsection*{An illustrative split step}

We include a brief description of a typical split step to illustrate RJMCMC. In the split step, we select an existing source $j^*$ at random and consider splitting it in two. In this case, $\thtkpp$ is the triplet $\{\z_{j^*},\w_{j^*},\s_{j^*}\}$. We draw random variables $\{\vect{r}_z,r_w,r_s\}$ from uniform distributions on appropriate intervals with zero mean, which to create two new sources $j^{*+}$ and $j^{*-}$ to replace $j^*$. The new source parameters are given by:

\begin{eqnarray*} \label{E:SplStp1}
\z_{j*\pm} &=& \z_{j*} \pm \vect{r}_z, \ \vect{r}_z(1) \sim U([-\frac{E_{z1}}{2},\frac{E_{z1}}{2}]), \ \vect{r}_z(2) \sim U([-\frac{E_{z2}}{2},\frac{E_{z2}}{2}])\\ \nonumber
w_{j*\pm} &=& w_{j*} \pm r_w, \ r_w \sim U([-E_{w}/2,E_{w}/2])\\ \nonumber
s_{j*\pm} &=& s_{j*} \pm r_s, \ r_s \sim U([-E_{s}/2,E_{s}/2])\\ \nonumber
\end{eqnarray*}

In this case, the Jacobian of the transformation (from \eqref{E:MtrHst2}) is equal to $2 \times 2 \times 2 = 8$. If we assume that source location components, widths and emission rates have independent uniform priors on $[0,R_{z1}]$, $[0,R_{z2}]$, $[0,R_{w}]$ and $[0,R_{s}]$, expressions for $f(\thtkpp' | \D, \thtnotkpp)$ and $f(\thtkpp | \D, \thtnotkpp)$ emerge:

\begin{equation*} \label{E:SplStp2}
f(\thtkpp | \D, \thtnotkpp) \sim f(\D | \thtkpp, \thtnotkpp) \times \frac{1}{R_{z1}R_{z2}R_{w}R_{s}} \times f(\text{Prior for all other parameters})
\end{equation*}

\begin{equation*} \label{E:SplStp3}
f(\thtkpp' | \D, \thtnotkpp) \sim f(\D | \thtkpp', \thtnotkpp) \times (\frac{1}{R_{z1}R_{z2}R_{w}R_{s}})^2 \times f(\text{Prior for all other parameters})
\end{equation*}

Noting that $g=(E_{z1}E_{z2}E_{w}E_{s})^{-1}$ and that $g'=1$, using \eqref{E:MtrHst2} we obtain the acceptance probability for splitting source $j^*$:

\begin{equation*} \label{E:MtrHstSpl}
\alpha(\text{split source $j^*$} | \thtnotkpp) = \max \left\{1,  \frac{f(\thtkpp' | \D, \thtnotkpp)}{f(\thtkpp | \D, \thtnotkpp)} \times \frac{j(\thtkpp'|\thtnotkpp)}{j(\thtkpp|\thtnotkpp)} \times \frac{E_{z1}E_{z2}E_{w}E_{s}}{R_{z1}R_{z2}R_{w}R_{s}} \times 8 \right\}
\end{equation*}

in which jump probabilities $j(\tht|\thtnotkpp)$ and $j(\tht'|\thtnotkpp)$ are equal.

\subsection{MCMC procedure}

A new state of the Markov chain having the posterior distribution of model parameters given observed data as stationary distribution, is created from the current state of the chain using the following procedure:

\begin{description}
\item[Update source parameters, $\{\z, \w, \s\}$.]
Firstly, source locations are updated sequentially using a Gaussian random-walk. Candidate locations outside the spatial domain of interest are rejected. Secondly, source widths are updated sequentially using a Gaussian random-walk. Finally, source emission rates are updated sequentially using a Gaussian random-walk.

\item[Update measurement error, $\sigma_Y$.]
Measurement error variance is updated using a Gaussian random walk on $\log_e \sigma_Y$.

\item[Update background parameters, $\bta$.]
The background parameter vector $\bta$ is updated by sampling from its full conditional.

\item[Optionally update wind direction bias, and plume opening angles, $\omega_H$, $\omega_V$.]
The bias in wind direction is updated using a Gaussian random walk \emph{modulo $2 \pi$}. Then the plume opening angles are updated if required using a Gaussian random walk on the log scale.

\item[Update number of sources, $m$.]
With equal probabilities of $0.25$, one of $4$ candidate states is generated, which if accepted, either increases (by birth or splitting) or decreases (by coalescence or death) the number of sources by $1$. Steps involving new states in which $m<0$ or $m>m_{\max}$ are rejected, where $m_{\max}$ is an upper bound for the number of sources.
\end{description}
%%%%%%%%%%%%%%%%%%%%%%%%%%%%%%%%%%%%%%%%%%%%%%%%%%%%%%%%%%%%

\end{appendices}

%%%%%%%%%%%%%%%%%%%%%%%%%%%%%%%%%%%%%%%%%%%%%%%%%%%%%%%%%%%%
%Table of contents
%\addcontentsline{toc}{chapter}{\numberline{}Bibliography}
%%%%%%%%%%%%%%%%%%%%%%%%%%%%%%%%%%%%%%%%%%%%%%%%%%%%%%%%%%%%

%%%%%%%%%%%%%%%%%%%%%%%%%%%%%%%%%%%%%%%%%%%%%%%%%%%%%%%%%%%%
%Bibliography
\pagebreak
%\section*{References}
\bibliography{LightTouch}
%\bibliography{AtmEnvMain_Arxiv}
%%%%%%%%%%%%%%%%%%%%%%%%%%%%%%%%%%%%%%%%%%%%%%%%%%%%%%%%%%%%
%\includepdf[pages={32,33}]{LTReport.pdf}
%%%%%%%%%%%%%%%%%%%%%%%%%%%%%%%%%%%%%%%%%%%%%%%%%%%%%%%%%%%%
%End document
\end{document}